\documentclass[11pt, reqno]{amsart}
\usepackage{graphicx}
\usepackage{xcolor}
\usepackage{bbm}
\usepackage[english]{babel}
\usepackage{empheq}
\usepackage{amssymb}
\usepackage[cp1252]{inputenc}
\usepackage{xfrac}

\numberwithin{equation}{section}

\newtheorem{definition}{Definition}

\newtheorem{theorem}{Theorem}
\newtheorem{proposition}{Proposition}
\newtheorem{corollary}{Corollary}
\newtheorem{lemma}{Lemma}
\newtheorem{remark}{Remark}

\begin{document}

\allowdisplaybreaks

\title[Optimal execution with dynamic risk adjustment]{Optimal execution with dynamic risk adjustment}
\author[X. Cheng, M. Di Giacinto and T.-H. Wang]{Xue Cheng, Marina Di Giacinto and Tai-Ho Wang}
\address{Xue Cheng \\
Department of Mathematical Finance\\
Peking University \\
Beijing, China}
\email{chengxue@math.pku.edu.cn}
\address{Marina Di Giacinto \\
Dipartimento di Economia e Giurisprudenza \\
Universit{\`a} degli studi di Cassino e del Lazio Meridionale, Cassino (FR), Italy \\
\emph{\textrm{and}} \\
Dipartimento di Matematica per le Scienze economiche, finanziarie ed attuariali \\
Universit{\`a} Cattolica del Sacro Cuore, Milano, Italy}
\email{digiacinto@unicas.it}
\address{Tai-Ho Wang \\
Department of Mathematics \\
Baruch College, The City University of New York \\
1 Bernard Baruch Way, New York, NY10010}
\email{tai-ho.wang@baruch.cuny.edu}

\begin{abstract}
This paper considers the problem of optimal liquidation of a position in a risky security in a financial market, where price evolution are risky and trades have an impact on price as well as uncertainty in the filling orders. The problem is formulated as a continuous time stochastic optimal control problem aiming at maximizing a generalized risk-adjusted profit and loss function. The expression of the risk adjustment is derived from the general theory of dynamic risk measures and is selected in the class of $g$-conditional risk measures. The resulting theoretical framework is nonclassical since the target function depends on backward components. We show that, under a quadratic specification of the driver of a backward stochastic differential equation, it is possible to find a closed form solution and an explicit expression of the optimal liquidation policies. In this way it is immediate to quantify the impact of risk-adjustment on the profit and loss and on the expression of the optimal liquidation policies.
\end{abstract}

\maketitle

\allowdisplaybreaks

\section{Introduction}

Trading algorithms are nowadays widely spread among financial agents. They are typically used for the execution of large orders by brokers. Differently from standard inventory models, once that a given market move has been decided, the success of the market operation crucially depends on the strategy chosen to execute the orders in the market. In fact, in addition to price fluctuation risk, optimal execution must also take into account the so-called market impact effect, i.e., the feedback effect that the order execution may have on the execution price. Practitioners have developed interesting models to quantify price impact and its implications, the simplest being the optimization of the Volume Weighted Average Price (VWAP) during the execution. A general introduction to the modeling problems and empirical evidence on algorithmic and high frequency trading can be found in \cite{Cartea-et-al}. More sophisticate models to quantify and manage price impact include \cite{Bertsimas-Lo}, \cite{Almgren-Chriss} in a discrete time model and \cite{Almgren} as its continuous time variant, and transient impact models such as \cite{Obizhaeva-Wang}, \cite{Alfonsi-Schied}, and \cite{Gatheral}.

In order to restate the problem of optimal execution as a variational problem, a crucial decision without a simple solution is the definition of the objective function to be optimized. While it is obvious that a risk neutral trader is willing to maximize the final profit and loss, the choice of the risk function that quantifies the tradeoff between profit and loss and execution risk is far from unique. In fact, the risk quantification is based on the full liquidation path and thus is essentially dynamic in nature.

In this paper we rely on the theory of dynamic convex risk measures and formulate the optimal liquidation problem as a stochastic control problem where the riskiness of the optimal strategy is quantified by a $g$-conditional measure. In this way, borrowing results from the general theory, see, e.g., \cite{Barrieu-ElKaroui} the riskiness of the liquidation strategy is locally described by the driver of a backward stochastic differential equation (BSDE).

The resulting optimal stochastic control problem is more general than a classical one due to the involvement of $g$-expectation which introduces additional terms depending non linearly from the volatilities of the backward components. Remarkably, we verify that for linear-quadratic expressions of the driver, the solution of this highly nonlinear controlled problem can be reduced to the solution of a system of Riccati ordinary differential equations (ODEs) and solved in closed form.

The solution provides a closed form description of the liquidation policy and of the impact that the risk aversion of the agent has on it. The explicit expression of the optimal control is useful to quantify the impact of the dynamic risk adjustment of the liquidation policy and offers interesting new insights showing the great potential relevance of forward-backward optimal control strategies in financial applications. Remarkably the analysis of the resulting control policies highlight the critical interaction between the uncertainty of order fills and the minimization of the dynamic risk functional. This interaction introduces a new characteristic trading time scale depending both on the microstructural parameters in the model and on the risk aversion that radically modifies in a non-linear way the optimal scheduled liquidation program and the policy used to implement it. As a consequence, the resulting optimal policy differs substantially from those that have been analyzed in the literature that consider only a forward risk adjustment component.

Many of these strategies can be recovered as limiting cases within our general framework. In the \cite{Almgren-Chriss} framework, by applying the technique of integration by parts, one can show that the optimal strategy for a risk neutral trader is always the VWAP strategy, no matter what the driving noise is so long as it is a martingale. However, for transient impact models such as \cite{Obizhaeva-Wang}, \cite{Alfonsi-Schied}, and \cite{Perdoiu-Shakhet-Shreve}, many of the optimal strategies are U-shaped like modified VWAP strategies. Namely, block trades at the beginning and the end of the trading period and a VWAP in between. See, for instance, \cite[Table~1, p.~20]{Obizhaeva-Wang}.

On the other hand, for risk averse traders the choice of risk factors for penalization varies case by case for the sake of tractability. In the
classical \cite{Almgren-Chriss} model, the authors use the quadratic variation as the risk factor; whereas \cite{Gatheral-Schied} use
time-average VaR as a risk factor to penalize the P\&L or cost of trading in the determination of optimal strategies (in fact, the authors claimed the
same rationale applies to the use of general coherent risk measure, see Remark~2.2 on p.~356). More recently, in \cite{Brigo-DiGraziano}, the asset is assumed following a displaced diffusion process and the authors compare the optimal trading strategies with various risk functions such as Value at Risk (VaR) and Expected Shortfall (ES). The authors also propose a new risk function called Squared Asset Expectation (SAE) in order to test the robustness of the optimal trading strategies. In \cite{Cuoco-He-Isaenko}, the problem of optimal execution is formalized as an exponential utility maximization program where the trader is subject to a VaR constraint that has to be instantaneously satisfied by P\&L for the entire  execution period. This paper in a sense attempts to blend two risk limiting elements (risk averse utility + VaR limit) in one control problem. \cite{Cartea-Jaimungal-SIAM} and \cite{Bank-Voss} discuss a control problem with linear quadratic objective function close to the one treated in this paper with the crucial difference that they do not consider the risk adjustment that induces the backward component. Finally, it is worth mentioning that in a discrete time model \cite{Lin-Chen-Pena} obtain an optimal trading strategy that is time-consistent and deterministic by minimizing a dynamic coherent risk of the cost of trading.

The latter paper is the one closest in spirit to ours, in fact they also use a dynamic risk measure to quantify the riskiness of a strategy. On the other hand the present analysis relies on continuous time and on a more realistic modeling of the profit and loss function where uncertainty in order fills, i.e., the risk for an order to be filled either incompletely or in excess, is taken into account. In addition, while their treatment relies on a recursive discrete time approach, our formulation involves a continuous time stochastic control problem that is solvable in closed form.

In fact, our approach is grounded on the formulation of general time-consistent dynamic convex risk measures relying on notion of the $g$-expectations developed by Peng (see, e.g., \cite{Peng-2004}) and his collaborators in the 90's. See \cite{Barrieu-ElKaroui} and the references therein for more details on the relationship between dynamical risk measures, $g$-expectations, and $g$-conditional risk measures.

Optimal strategies with uncertain order fills have been introduced by \cite{Xue}, and, more recently, treated in \cite{Bulthuis-et-al} and \cite{Vaes-et-Hauser-arxiv}. In particular, in the \cite{Xue} framework the magnitude of the uncertain order fills is assumed to be proportional to the order size, while \cite{Vaes-et-Hauser-arxiv} develop a discrete time model showing that the volume uncertainty is independent of trader's decisions and  the optimal strategy for a risk-averse investor is a trade-off between early and late trades to balance risk associated to both price and volume.

Furthermore, our model provides for a quadratic running penalty to discourage large deviation from a given threshold chosen by the investment committee.

The use of a quadratic penalization for deviations of the trading velocity from a pre-specified target is also found in other classes of trading problems, such as hedging via risk minimization with constraints (see, \cite{Lee}) or preventing the agent from trading too quickly with either market or limit order type (see, \cite{Bulthuis-et-al}).

The paper is organized as follows. Section \ref{sect:the-model} sets up the model and provides more detailed discussions and motivations on the problem formulation. Section \ref{sect:dynamic-risk-measure} introduces the general notion of convex risk measures and the way these are used to \textquotedblleft risk adjust\textquotedblright\ the target functional form. Section \ref{sect:optimal-control-problem} discusses the stochastic control problem and its solution. In Section \ref{sect:opt-liquidation-strategies} we briefly analyze the applicative implications of the result and conclude.

\section{The model} \label{sect:the-model}

To setup the mathematical model, let $[0,T]$, $T > 0$, be the fixed finite trading horizon and let $\left(\Omega, \mathcal{F}, \mathbb{P}\right)$ be a complete probability space that carries a two-dimensional uncorrelated Brownian motion $\left(B_{1}, B_{2}\right) := \{B_{1}(t),B_{2}(t)\}_{t \in [0,T]}$. The information structure is described by the filtration $\mathbb{F} := \left\{\mathcal{F}_{t}\right\}_{t \in [0,T]}$ generated by the trajectories of the two-dimensional Brownian motion, and completed with the addition of the all $\mathbb{P}$-null measure sets of $\mathcal{F}$. We denote by $\mathcal{M}^{2}_{\mathbb{F}} (0,T; \mathbb{R}^{2})$ the set of all two-dimensional real-valued $\mathbb{F}$-predictable processes $\left\{H(t)\right\}_{t \in [0,T]}$ satisfying $\mathbb{E} \left[\int_{0}^{T}\left|H(t)\right|^{2}dt \right] < +\infty$, $\mathcal{H}^{2}_{\mathbb{F}} (0,T; \mathbb{R})$ the set of all real-valued $\mathbb{F}$-progressively measurable processes $\left\{H(t)\right\}_{t \in [0,T]}$ satisfying $\mathbb{E}\left[\int_{0}^{T}\left|H(t)\right|^{2}dt \right] < +\infty$, $L^{2}_{\mathcal{A}} (\Omega; \mathbb{R})$ the set of all real-valued $\mathcal{A}$-measurable square integrable random variables, $L^{\infty}_{\mathcal{A}} (\Omega; \mathbb{R})$ the set of all bounded real-valued $\mathcal{A}$-measurable random variables, where $\mathcal{A} \subseteq \mathcal{F}$ is a sub-$\sigma$-algebra, and $\mathcal{S}^{n}$ the set of all $n\times n$ symmetric real matrices.

An issue often overlooked in the existing literature on optimal execution is that the pre-scheduled transactions might be underly or overly executed. As discussed in \cite{Xue}, in modern electronic markets order execution is a complex process involving potentially different type of orders and realized transaction may deviate from scheduled ones. We refer to such a deviation as the uncertainty of order fills. To take into consideration this kind of risk, we add a noise term to the evolution of the investor's position $X$ and therefore suppose that it satisfies the following stochastic differential equation:
\begin{equation} \label{eq:position}
\begin{cases}
dX(t)=-v(t)dt+mdB_{1}(t),\quad t\in [0,T], \\
X(0)=x_{0}>0,
\end{cases}
\end{equation}
where the $\mathbb{F}$-progressively measurable process $v\colon \Omega \times[0,T]\rightarrow \mathbb{R}$ is the trading rate of market order and regarded as the control variable, while $m\geq 0$ measures the magnitude of the uncertainty of order fills.

Modeling the uncertainty of order fills by a diffusion process driven by a Brownian motion serves as a `zeroth order approximation' to the uncertainty and the implications seem plausible as far as the size of the uncertainty is small and $m \rightarrow 0$. A more accurate model for the uncertainty of order fills is by adding a pure jump spectrally negative Lévy process. Alternatively, the term $m dB_{1}$ can be also interpreted as the whole retail of the financial institution exposure. In other words, the initial amount $x_{0}$ to be liquidated from the responsible of the desk of the whole financial institution is subject to small random variation induced by additional liquidation or redemption orders that he could receive from other desks in the meantime.

The fair price $S$ of the stock follows the dynamics:
\begin{equation*}
\begin{cases}
dS(t)=\gamma dX(t)+\sigma dB_{2}(t),\quad t\in [0,T], \\
S(0)=s_{0}>0,
\end{cases}
\end{equation*}
i.e.,
\begin{equation*}
\begin{cases}
dS(t)=-\gamma v(t)dt+\gamma mdB_{1}(t)+\sigma dB_{2}(t),\quad t\in [ 0,T], \\
S(0)=s_{0}>0.
\end{cases}
\end{equation*}
In other words, the fair price $S$ is driven by an Arithmetic Brownian motion with drift equal to zero and volatility $\sigma >0$, along with a linear permanent impact with parameter $\gamma \geq 0$. The choice of considering a Bachelier price evolution (see, e.g., \cite{Bachelier,Courtault-et-al}) is made to keep the problem tractable. It restricts the applicability of the model to securities-trading in a low volatility environment (quite common in recent years).

Following \cite{Almgren-Chriss}, the transacted price $\widetilde{S}$ is consists of the fair price and a slippage referred as temporary impact:
\begin{equation} \label{eq:tp}
\widetilde{S}(t)=S(t)-\eta v(t),\quad t\in [ 0,T].
\end{equation}
That is, the transacted price reflects a temporary impact given by a linear function of the current trading rate of market order $v$ with size $\eta >0$.

Following \cite[Section 2.4, p.~10]{Almgren-Chriss}, the profit and loss (P\&L) $\Pi^{0}(t)$ of a trading strategy earned over the time interval $[0,t]$, $t\leq T$, is defined as:
\begin{equation}\label{eq:P&L}
\Pi^{0}(t) := X(t)\left( S(t)-S(0)\right) +\int_{0}^{t} (S(0)-\widetilde{S}(u)) dX(u).
\end{equation}
The first term of the right-hand side captures the change in fair value of the remaining untransacted shares, while the second term measures the transaction costs resulting from selling shares due to the presence of a temporary price impact component. More details can be found in Appendix \ref{app:P&L}, Subsection \ref{app:decomposition-P&L}, p. \pageref{app:decomposition-P&L}.

Taking into account \eqref{eq:position}--\eqref{eq:tp}, the computation given in Appendix \ref{app:P&L} (Subsection \ref{app:computation-P&L}, p. \pageref{app:computation-P&L}) shows that:
\begin{multline*}
\Pi^{0}(t) = \frac{\gamma}{2} \left(X^{2}(t) - x^{2}_{0}\right) + \frac{\gamma}{2} m^{2} t -\eta\int_{0}^{t} v^{2}(u)du + \\
+ \eta m\int_{0}^{t} v(u) dB_{1}(u) + \sigma \int_{0}^{t} X(u) dB_{2}(u),
\end{multline*}
or, equivalently:
\begin{multline*}
\Pi^{0}(t) = \gamma m^{2} t -\int_{0}^{t}\left( \gamma v(u)X(u) + \eta v^{2}(u)\right) du + \\
+ m\int_{0}^{t} \left(\gamma X(u) + \eta v(u) \right) dB_{1}(u) + \sigma \int_{0}^{t} X(u) dB_{2}(u).
\end{multline*}

Depending on the circumstances, we will use indifferently the first or second expression.

We must ensure that at $t=T$ the initial position is fully liquidated despite the uncertainty of order fills. To this end, we add to the P\&L at the terminal time $T$ a penalty term for the final block trade $f\colon \mathbb{R} \rightarrow [0, +\infty)$ so that anything but complete liquidation is undesirable. Following \cite{Xue}, we take into consideration as a penalty term for the final block trade $x \in \mathbb{R}$ a continuous function defined as:
\begin{equation} \label{eq:penalty-function}
f(x) := \beta x^{2}, \qquad \beta > 0.
\end{equation}

Observe that any final block trade is discouraged since it is zero if and only if the initial position is fully liquidated; otherwise it is always positive if the initial position is underly or overly liquidated. Notice that this choice of the terminal condition penalizes both negative or positive final investor's position. This is consistent with the task to be accomplished by the desk that is in charge of the full portfolio liquidation.

Finally, we introduce a running penalty $h\colon \mathbb{R}\rightarrow [0,+\infty )$ with the following quadratic specification:
\begin{equation} \label{eq:running-penalty}
h(v(t)):=\lambda_{1}\left( v(t)-\overline{v}\right)^{2},\quad t \in [0,T].
\end{equation}
where $\lambda_{1}\geq 0$ is the cost to be paid for a unit deviation per unit of time from the desired target $\overline{v}>0$. This target speed of execution represents the ideal liquidation rate which is exogenously set by the investment committee. Note that the above function penalizes during liquidation any deviation of the selected trading rate from $\overline{v}$ and prevents either overly large trading rates or the placement of buy side orders.

\section{Risk adjustment of the target profit and loss function} \label{sect:dynamic-risk-measure}

It is a well known result of financial economics that in a static single period market model, the preference of a risk averse non-satiated agent can be represented using a concave and increasing functional, a utility function. In fact, it is easy to verify that the degree of concavity of the utility function is proportional to the additional amount of money (risk compensation) in addition to the expected payoff that the agent requires to play a fair lottery. The quantification of this risk compensation is critical to interpret the best risk-return tradeoff achievable in the market and thus sort out the best investment opportunities.

A similar dynamic risk-return tradeoff is faced by investors that are willing to liquidate their portfolios in the market and are subject to the uncertainty induced by price impact and by limited liquidation possibilities due to the microstructural frictions that typically affect real markets. While conventional approaches to optimal liquidation maximize the profit and loss function that accounts for the costs and profits generated by the execution problem, the main innovation of this paper is the formulation and the solution of an optimization program that evaluates the liquidation policy by taking into account also the risk aversion of the investor. In other words, we specify a functional that represents both the cost-benefit tradeoff of a liquidation strategy while penalizing liquidation paths that are particularly risky from the point of view of the investor. A natural way to include this risk contribution in a dynamic functional to be optimized is to rely on the theory of $g$-conditional risk measures, i.e., dynamic risk measures characterized by the solution to BSDEs associated with a convex driver $g$.

In Appendix \ref{app:risk-measures} we briefly review the basic definitions and results on dynamic risk measures that are relevant to our analysis.

We will focus our analysis to the case of the so-called dynamic entropic risk measure defined, for any $\xi (T)\in L_{\mathcal{F}_{T}}^{2}(\Omega; \mathbb{R})$, as follows:
\begin{equation*}
\mathcal{R}\left( t,\xi (T)\right) :=\frac{1}{\lambda_{2}} \ln\mathbb{E} \left[ \exp \left( -\lambda_{2}\xi (T)\right) \mid \mathcal{F}_{t}\right], \quad t \in [0,T]
\end{equation*}
where $\lambda_{2}\geq 0$ is the risk aversion coefficient, that is, everything else equal, a change of $\lambda_{2}$ modifies the risk attitude of the investor. The next Proposition establishes a well-known connection between the entropic risk measure and the solution of the one-dimensional BSDE with the following quadratic driver $g$:
\begin{equation*}
g\left( Z_{1}(t),Z_{2}(t)\right) :=\dfrac{\lambda_{2}}{2}\left(Z_{1}^{2}(t)+Z_{2}^{2}(t)\right) , \quad t \in [0,T]
\end{equation*}
where $\left(Z_{1},Z_{2}\right)^{\top} := \{ \left(Z_{1}, Z_{2}\right)^{\top} \}_{t\in [0,T]}$ is a two-dimensional BSDE control process corresponding to the two dimensional correlated Brownian motion $\left(B_{1},B_{2}\right)$.

\begin{proposition} \label{prop:erm}
Let $\left(Z_{1}, Z_{2}\right)^{\top} \in \mathcal{M}_{\mathbb{F}}^{2}(0,T; \mathbb{R}^{2})$. For any $\xi (T) \in L^{2}_{\mathcal{F}_{T}} (\Omega; \mathbb{R})$, the entropic risk measure is the unique solution to the following one-dimensional BSDE:
\begin{equation} \label{eq:BSDE-entropic}
\begin{cases}
\begin{aligned} d\mathcal{R}\left(t, \xi (T)\right) = &-\dfrac{\lambda_{2}}{2} \left(Z^{2}_{1}(t) + Z^{2}_{2}(t)\right) dt \\
&+Z_{1}(t)dB_{1}(t) +Z_{2}(t)dB_{2}(t), \quad t \in [0,T],
\end{aligned} \\
\mathcal{R}(T, \xi (T)) =-\xi (T).
\end{cases}
\end{equation}
\end{proposition}

\begin{proof}
See Appendix \ref{app:proofs}, p. \pageref{app:erm}.
\end{proof}

\begin{remark} \label{rmk:sg-property}
For any time $t\in [0, T]$ and $\tau \in [t, T]$, $\mathcal{R}$ corresponds to the solution flow to the above BSDE \eqref{eq:BSDE-entropic} and has the following semigroup property:\footnote{The relation is verified in Appendix \ref{app:proofs}, p. \pageref{app:sg-property}.}
\begin{equation*}
\mathcal{R}\left( t,\mathcal{R}(\tau, \xi (T))\right) =\mathcal{R} \left(t,\xi (T)\right) \quad \mathbb{P}\text{-a.s.}.
\end{equation*}
\end{remark}

The same type of relationship can be extended to a larger class of dynamic convex risk measures by considering a more general expression for the driver
of a BSDE (see Appendix \ref{app:risk-measures}, Proposition \ref{th:risk-measure}).

These results show that the driver of a BSDE is a natural object to describe locally (i.e., over small intervals of time) the dynamic risk-return tradeoff faced by an investor who liquidates her position in the market and measures the risk considering a $g$-conditional risk measure.

In our framework, an important motivation leading us to the use of a dynamic risk measure arises from the fact that the success of the trader's liquidation policy is based on the ability of two countervailing tasks: on one hand, it is related to her ability to complete the liquidation of the portfolio with a maximum profit (equivalently minimum loss) by the final date; on the other hand, the liquidation policy is conditioned to the unfolding of the uncertainty drivers that impact both on prices and quantities. At each time, the dynamic risk assessment of the final cost to be paid takes into account both the costs of final partial liquidation and the evaluation of this potential cost in a way that is conditional on the information available and adjusted to keep into account the risk aversion of the trader.

\section{The optimal control problem} \label{sect:optimal-control-problem}

The optimal execution is formulated and studied as a stochastic optimal control problem. For any initial time $t \in[0,T]$ and initial points $X(t)
:= x \in \mathbb{R}$, the trader maximizes her expected risk adjusted total P\&L of liquidation within the time horizon $[t,T]$, penalized by the final block trade and by the cumulative deviation from the pre-specified target speed $\overline{v}$.

\subsection{The state equation and the objective functional}

Let $\mathcal{F}^{t}_{u}$ be the $\sigma$-algebra generated by $\left\{B_{1}(r)-B_{1}(t), B_{2}(r)-B_{2}(t)\right\}_{r\in[t,u]}$ and $\mathbb{F}^{t} := \left\{ \mathcal{F}^{t}_{u} \right\}_{u \in [t,T]}$ the filtration augmented by all $\mathbb{P}$-null measure sets of $\mathcal{F}$. In order to write up the state equation and the objective functional, let $\Pi^{t}(T)$ denote the P\&L earned over the time interval $[t,T] $, i.e.,
\begin{equation*}
\Pi^{t}(T):=\Pi^{0}(T)-\Pi^{0}(t).
\end{equation*}

Setting:
\begin{equation*}
\begin{cases}
-dY(u):=d\Pi ^{u}(T)-d\mathcal{R}(u,f(X(T)))-d\left(\int_{u}^{T}h(v(u))du\right) ,\quad u\in [t,T], \\
Y(T):=-f(X(T))=-\beta X^{2}(T),
\end{cases}
\end{equation*}
where $f$ is the cost function given by \eqref{eq:penalty-function} and $h$ is the running penalty defined in \eqref{eq:running-penalty}, the state equation is described by the following decoupled quadratic growth forward-backward stochastic differential equation (qgFBSDE):
\begin{equation} \label{eq:qgFBSDE}
\begin{cases}
dX(u)=-v(u)du+mdB_{1}(u),\quad u\in [t,T], \\
\begin{aligned}
-dY(u) = &\;\widetilde{g}(X(u), \widetilde{Z}_{1}(u), \widetilde{Z}_{2}(u), v(u))du + \\ &-\widetilde{Z}_{1}(u)dB_{1}(u) -\widetilde{Z}_{2}(u)dB_{2}(u), \quad u \in [t,T],
\end{aligned} \\
X(t)=x, \qquad Y(T)=-\beta X^{2}(T),
\end{cases}
\end{equation}
with
\begin{multline} \label{eq:gtilde}
\widetilde{g}(X(u),\widetilde{Z}_{1}(u),\widetilde{Z}_{2}(u),v(u)):=\left( \eta +\lambda _{1}\right) v^{2}(t)+\dfrac{\lambda_{2}}{2} (\widetilde{Z}_{1}^{2}(u)+\widetilde{Z}_{2}^{2}(u))+   \\
+\left( \gamma X(u)-2\lambda _{1}\overline{v}\right) v(u)-\left( \gamma m^{2}-\lambda _{1}\overline{v}^{2}\right) ,\quad u\in [t,T],
\end{multline}
and
\begin{equation} \label{eq:Ztilde}
\left(
\begin{matrix}
\widetilde{Z}_{1} \\
\widetilde{Z}_{2}
\end{matrix}
\right) (u)=\left(
\begin{matrix}
Z_{1}(u)+\gamma mX(u)+m\eta v(u) \\
Z_{2}(u)+\sigma X(u)
\end{matrix}
\right) ,\quad u\in [t,T].
\end{equation}

\begin{remark}
In this framework, the terminal condition in \eqref{eq:qgFBSDE} allows us to measure at any time the riskiness of the final penalty.
\end{remark}

Here, the set of admissible control is a space of processes defined as follow:
\begin{equation*}
\mathcal{V}_{\text{ad}}[t,T]:=\left\{ v:[t,T]\times \Omega \rightarrow \mathbb{R}\,|\,v\in \mathcal{H}_{\mathbb{F}^{t}}^{2}(t,T;\mathbb{R})\right\}.
\end{equation*}

Observe that for any, $v(\cdot) \in \mathcal{V}_{\text{ad}}[T,t]$, the above decoupled qgFBSDE \eqref{eq:qgFBSDE} admits a unique strong solution. As a matter of fact, existence and uniqueness of the solution to the forward component is a rather standard result (see, e.g., \cite[Theorem~6.3, p.~42]{Yong-Zhou}); while the solution to the BSDE with quadratic growth driver and unbounded terminal condition in \eqref{eq:qgFBSDE} is guaranteed by, e.g., \cite{Briand-Hu2}.

Denoting by $( Y, \widetilde{Z}_{1}, \widetilde{Z}_{2})^{\top} (\cdot; t,x,v(\cdot))$ the solution to the backward component of \eqref{eq:qgFBSDE}, when $X(\cdot; t,x,v(\cdot))$ is the solution to the forward part of \eqref{eq:qgFBSDE} starting from $x \in \mathbb{R}$ at time $t \in [0,T]$ and control $v(\cdot) \in \mathcal{V}_{\text{ad}}[t,T]$, the objective functional is given by:
\begin{equation} \label{eq:functional}
J(t,x; v(\cdot)) := Y(t;t,x;v(\cdot)),
\end{equation}
and the trader's optimal liquidation policy consists in finding for any $(t,x)\in [0,T] \times \mathbb{R}$ the solution to the following problem:
\begin{equation} \label{eq:opt-problem}
\textsc{maximize}\quad J(t,x; v(\cdot))\quad \textsc{over}\quad v(\cdot) \in \mathcal{V}_{\text{ad}}[t,T],
\end{equation}
while the associated value function $W$ is thus defined as:
\begin{equation*}
\begin{cases}
\displaystyle{W(t,x) := \sup_{v(\cdot )\in \mathcal{V}_{\textsl{ad}}[t,T]} J(t,x;v(\cdot )), \quad \forall (t,x)\in [ 0,T]\times \mathbb{R}^{2}} \\
W(T,x):=-\beta x^{2},\quad \forall x\in \mathbb{R}.
\end{cases}
\end{equation*}

Notice that the expression of $\mathcal{R}(t,f(X(T)))$ depends on the backward component and makes the full stochastic optimal control problem non-standard.

\subsection{The HJB equation}

In the context of stochastic optimal control problems with finite horizon, it is well-known that the value function is associated to a second-order partial differential equation (PDE) with terminal boundary condition -- the so-called Hamilton-Jacobi-Bellman (HJB) equation -- which we aim to derive.

Following, e.g., \cite{Peng-1997}, it is possible to derive the generalized HJB equation associated with the controlled state equation \eqref{eq:qgFBSDE}. It is given by:
\begin{equation} \label{eq:HJB}
\begin{cases}
w_{t}(t,x)+\displaystyle{\sup_{v\in \mathbb{R}}} \, \mathcal{H}_{cv} \left( x, w_{x}, w_{xx}; v\right) = 0, \quad (t,x)\in [0,T]\times \mathbb{R}, \\
w(T,x)=-\beta x^{2},
\end{cases}
\end{equation}
where the Hamiltonian current value $\mathcal{H}_{cv}$ reads as:
\begin{multline} \label{eq:H-cv}
\mathcal{H}_{cv}\left( x,q,Q;v\right) := \operatorname{tr}[\Sigma \Sigma^{\top} \mathrm{Q}]+\langle b(v),q\rangle -\widetilde{g}(x,\Sigma^{\top }q,v), \\
(x,q,Q) \in \mathbb{R} \times \mathbb{R} \times \mathbb{R} \times \mathcal{S}^{2},
\end{multline}
with $\Sigma$ and $b \colon \mathbb{R} \rightarrow \mathbb{R}^{2}$ which represent the volatility matrix and the drift term of the forward diffusion process in \eqref{eq:qgFBSDE}, respectively, i.e.,
\begin{equation} \label{eq:vol-matrix}
\Sigma := \left(
\begin{matrix}
m & 0 \\
\gamma m & \sigma
\end{matrix}
\right), \qquad b(v):=\left(
\begin{matrix}
-v \\
-\gamma v
\end{matrix}
\right),
\end{equation}
and $\widetilde{g}$ is specified in \eqref{eq:gtilde}.

The construction of the above generalized HJB \eqref{eq:HJB} follows repeating the well-known argument applied to state the HJB equation in standard control problems. The additional prescription that the differential representation of the stochastic backward variable $\left(Z_{1},Z_{2}\right)^{\top}$ is given by $\Sigma^{\top }Dw$, follows from the standard Feynman-Kac representations for FBSDEs. Optimality for the solution to the generalized HJB \eqref{eq:HJB} for the original control problem will be proved in the verification argument.

The function $\mathcal{H}_{cv}$ has a unique maximum point on $\mathbb{R}$ given by:
\begin{equation} \label{eq:maxpoint}
v^{\star}(x,q,Q) = -\frac{q +\gamma x -2\lambda_{1} \overline{v}}{2\left(\eta +\lambda_{1}\right)}.
\end{equation}

Therefore, the Hamilton-Jacobi-Bellman equation related to the stochastic control \eqref{eq:opt-problem} can be rewritten as:
\begin{equation} \label{eq:HJB-scalar}
w_{t} + \frac{1}{2}m^{2}w_{xx} -\frac{1}{2}\lambda_{2} m^{2}w_{x}^{2} + \gamma m^{2} -\lambda_{1} \overline{v}^{2} + \frac{1}{4\left(\eta +\lambda_{1}\right)}\left(w_{x} +\gamma x -2\lambda_{1}\overline{v} \right)^{2} =0,
\end{equation}
with terminal condition:
\begin{equation} \label{eq:HJB-tc}
w(T,x) = -\beta x^{2}.
\end{equation}

\subsection{Solution to the HJB equation and the verification theorem}

The value function and the optimal trading strategy are presented in the verification theorem below. In order to state and prove the result, we start with the following lemma making use for convenience of the following notation:
\begin{equation} \label{eq:rho}
\kappa:= 2m^{2}\left(\eta +\lambda_{1}\right).
\end{equation}

\begin{lemma} \label{lemma:HJB-solution}
Let $\beta > \dfrac{\gamma}{2}$ and $\lambda_{2} < \dfrac{1}\kappa$. Then the deterministic functions $a,b,c \colon [0,T] \rightarrow \mathbb{R}$ uniquely solve the following system of Riccati ODEs:
\begin{equation*}
\begin{cases}
\dot{a}(t) = \left[m^{2}\lambda_{2} - \dfrac{1}{2\left(\eta + \lambda_{1}\right)}\right] a^{2}(t) -2m^{2}\lambda_{2} \gamma a(t) + m^{2}\lambda_{2} \gamma^{2}, \quad a(T) = -2\beta +\gamma, \\[9pt]
\dot{b}(t) = \left[m^{2}\lambda_{2} - \dfrac{1}{2\left(\eta + \lambda_{1}\right)}\right] a(t)b(t) -m^{2}\lambda_{2} \gamma b(t) + \dfrac{\lambda_{1}\overline{v}}{\eta + \lambda_{1}} a(t), \quad b(T)=0 , \\[9pt]
\begin{aligned} \dot{c}(t) = &\dfrac{1}{2} \left[\lambda_{2} m^{2} -\dfrac{1}{2\left(\eta + \lambda_{1}\right)}\right] b^{2}(t) +\dfrac{\lambda_{1}\overline{v}}{\eta + \lambda_{1}} b(t) -\dfrac{1}{2}m^{2} a(t) +\\
&-\frac{1}{2}m^{2}\gamma +\lambda_{1} \overline{v}^{2} -\dfrac{\lambda_{1}^{2} \overline{v}^{2}}{\eta +\lambda_{1}}, \quad c(T) = 0,
\end{aligned}
\end{cases}
\end{equation*}
i.e.,
\begin{equation} \label{eq:ODEs-solution}
\begin{cases}
a(t)= -\gamma \sqrt{\kappa\lambda_{2}} \frac{2\beta \sqrt{\kappa\lambda_{2}} \sinh \left[ \frac{\gamma \sqrt{\kappa \lambda_{2}}}{2\left(\eta +\lambda_{1}\right)} (T-t)\right] + \left(2\beta -\gamma\right) \cosh \left[ \frac{\gamma \sqrt{\kappa \lambda_{2}}}{2\left(\eta +\lambda_{1}\right)}(T-t) \right]}{\left[2\beta \left(1 -\kappa\lambda_{2} \right) -\gamma\right] \sinh \left[ \frac{\gamma \sqrt{\kappa \lambda_{2}}}{2\left(\eta +\lambda_{1}\right)} (T-t)\right] +\gamma \sqrt{\kappa\lambda_{2}} \cosh \left[ \frac{\gamma \sqrt{\kappa \lambda_{2}}}{2\left(\eta +\lambda_{1}\right)} (T-t)\right]} \\[15pt]
b(t) = 2\lambda_{1}\overline{v}\frac{(2\beta -\gamma)\sinh \left[ \frac{\gamma \sqrt{\kappa \lambda_{2}}}{2\left(\eta +\lambda_{1}\right)} (T-t) \right] +2\beta \sqrt{\kappa\lambda_{2}} \left[ \cosh\left[\frac{\gamma \sqrt{\kappa \lambda_{2}}}{2\left(\eta +\lambda_{1}\right)} (T-t)\right] -1 \right]}{\left[ 2\beta \left(1 - \kappa\lambda_{2}\right) -\gamma \right] \sinh \left[ \frac{\gamma \sqrt{\kappa \lambda_{2}}}{2\left(\eta +\lambda_{1}\right)} (T-t)\right] +\gamma \sqrt{\kappa\lambda_{2}} \cosh \left[ \frac{\gamma \sqrt{\kappa \lambda_{2}}}{2\left(\eta +\lambda_{1}\right)} (T-t)\right]} \\[15pt]
\begin{array}{ll}
c(t)= \hspace{-9pt} & -\frac{8\beta \lambda_{1}^{2} \overline{v}^{2}}{\gamma^{2}} + \left[ m^{2}\gamma -\frac{1}{2} \frac{\gamma m^{2}}{1 -\kappa\lambda_{2}} -\lambda_{1} \overline{v}^{2}\right] (T-t)+ \\
& +\frac{2\lambda_{1}^{2}\overline{v}^{2}}{\gamma^{2}} \frac{4\gamma \beta \sqrt{\kappa\lambda_{2}} + \left[(2\beta -\gamma)^{2} + 4\beta^{2}\left( 1- 2\kappa\lambda_{2} \right) \right] \sinh \left[ \frac{\gamma \sqrt{\kappa \lambda_{2}}}{2\left(\eta +\lambda_{1}\right)} (T-t) \right]}{\left[ 2\beta\left(1 -\kappa\lambda_{2} \right) -\gamma\right] \sinh \left[ \frac{\gamma \sqrt{\kappa \lambda_{2}}}{2\left(\eta +\lambda_{1}\right)} (T-t) \right] +\gamma \sqrt{\kappa\lambda_{2}}\cosh \left[ \frac{\gamma \sqrt{\kappa \lambda_{2}}}{2\left(\eta +\lambda_{1}\right)} (T-t) \right]} + \\[12pt]
& +\frac{m^{2}\left(\eta + \lambda_{1}\right)}{1- \kappa\lambda_{2}} \ln \left[ \left\vert \frac{2\beta \left(1- \kappa\lambda_{2}\right) - \gamma}{\gamma \sqrt{\kappa\lambda_{2}}} \mbox{\footnotesize $\sinh \left[ \frac{\gamma \sqrt{\kappa \lambda_{2}}}{2\left(\eta +\lambda_{1}\right)} (T-t) \right] + \cosh \left[ \frac{\gamma \sqrt{\kappa \lambda_{2}}}{2\left(\eta +\lambda_{1}\right)} (T-t) \right]$} \right\vert \right]
\end{array}
\end{cases}
\end{equation}
and the following concave function:
\begin{equation} \label{eq:ansatz}
w(t,x) = \frac{1}{2}\left(a(t)-\gamma\right) x^2 + b(t)x + c(t), \quad t \in [0,T],
\end{equation}
satisfies the HJB equation \eqref{eq:HJB-scalar}--\eqref{eq:HJB-tc} for $x \in \mathbb{R}$.
\end{lemma}

\begin{proof}
See Appendix \ref{app:proofs}, p. \pageref{app:HJB-solution}.
\end{proof}

Taking into account \eqref{eq:maxpoint} and Lemma \ref{lemma:HJB-solution}, the feedback map coming from the optimization of the Hamiltonian current value $\mathcal{H}_{cv}$ defined in \eqref{eq:H-cv} when $\operatorname{D}w$ is plugged in place the formal argument $q$ reads as:
\begin{equation*}
(t,x) \longmapsto v^{\star}(t,x) := -\frac{a(t)x +b(t)-2\lambda_{1} \overline{v}}{2\left(\eta + \lambda_{1}\right)},
\end{equation*}
and applying \eqref{eq:Ztilde} along with Lemma \ref{lemma:HJB-solution} the backward control process in state feedback form is given by:
\begin{equation*}
(t,x) \longmapsto \left(
\begin{matrix}
\widetilde{Z}^{\star}_{1} \\
\widetilde{Z}^{\star}_{2}
\end{matrix}
\right)(t,x) := \left(
\begin{matrix}
-m \left[\left(a(t) -\gamma\right)x - b(t)\right] \\
0
\end{matrix}
\right).
\end{equation*}

Thus, the corresponding closed loop equation:
\begin{equation} \label{eq:cle}
\begin{cases}
dX(u)=-v^{\star }(u,X(u))du+mdB_{1}(u), \quad u\in [ t,T], \\[6pt]
\begin{aligned}
dY(u) &= \widetilde{g}(X(u), \widetilde{Z}^{\star}_{1}(u), \widetilde{Z}^{\star}_{2}(u), v^{\star}(u,X(u)))du \,+ \\
&\quad-\widetilde{Z}^{\star}_{1}(u,X(u)) dB_{1}(u) -\widetilde{Z}^{\star}_{2}(u,X(u)) dB_{2}(u), \quad u \in [t,T],
\end{aligned} \\
X(t)=x, \qquad Y(T)=-\beta X^{2}(T),
\end{cases}
\end{equation}
with $\widetilde{g}$ given by \eqref{eq:gtilde}, has a unique solution\footnote{As previously mentioned, see, e.g., \cite[Theorem~6.3, p.~42]{Yong-Zhou} for the forward part and \cite{Briand-Hu2} for the backward component.}. Moreover, denoting by $X^{\star}(\cdot) := X^{\star} (\cdot; t,x,v^{\star}(\cdot))$ the solution to the forward process of the above closed loop equation \eqref{eq:cle}, the feedback strategy defined as:
\begin{equation*}
v^{\star }(u):=v^{\star }(u,X^{\star }(u)), \quad u \in [t,T],
\end{equation*}
is Lipschitz continuous. Therefore $v^{\star }(\cdot )$ is admissible, that is, $v^{\star }(\cdot )\in \mathcal{V}_{\text{ad}}[t,T]$.

Now we are ready to prove the following verification theorem.

\begin{theorem} \label{th:vt}
Let $\beta >\dfrac{\gamma }{2}$, $\lambda_{2}<\dfrac{1}{\kappa }$, and $w$ be the function defined in \eqref{eq:ansatz}. Then, for any $(t,x)\in [ 0,T]\times \mathbb{R}$, the optimization problem \eqref{eq:opt-problem} has a unique solution corresponding to the value function $W(t,x)=w(t,x)$, i.e.,
\begin{equation*}
W(t,x)=\frac{1}{2}\left( a(t)-\gamma \right) x^{2}+b(t)x+c(t).
\end{equation*}
The unique optimal scheduled trading rate $v^{\star }(\cdot )$ in state feedback form is given by:
\begin{equation} \label{eq:v-star}
v^{\star }(u)=-\frac{a(u)X^{\star }(u)+b(u)-2\lambda_{1} \overline{v}}{2\left( \eta +\lambda_{1}\right) }\quad \mathbb{P}\text{-a.s.}, \quad u\in [t,T],
\end{equation}
and the unique two-dimensional optimal backward control process $(\widetilde{Z}_{1}^{\star}, \widetilde{Z}_{2}^{\star})^{\top}$ in feedback form reads as:
\begin{equation} \label{eq:Ztilde-star}
\left(
\begin{matrix}
\widetilde{Z}_{1}^{\star } \\
\widetilde{Z}_{2}^{\star }
\end{matrix}
\right) (u)=\left(
\begin{matrix}
-m\left[ \left( a(u)-\gamma \right) X^{\star }(u)+b(u)\right] , \\
0
\end{matrix}
\right) \quad \mathbb{P}\text{-a.s.}, \quad u \in [t,T].
\end{equation}
\end{theorem}

\begin{proof}
See Appendix \ref{app:proofs}, p. \pageref{app:vt}.
\end{proof}

An immediate consequence of the above result is the following.

\begin{corollary} \label{cor:Z-star}
The unique two-dimensional optimal backward control process $\left( Z_{1}^{\star}, Z_{2}^{\star}\right)^{\top}$ in state feedback form related to the dynamic entropic risk measure \eqref{eq:BSDE-entropic} is given by:
\begin{equation*}
\left(
\begin{matrix}
Z_{1}^{\star} \\
Z_{2}^{\star}
\end{matrix}
\right) (u) = \left(
\begin{matrix}
-\dfrac{m}{2\left(\eta + \lambda_{1}\right)} \left[\left(\eta + 2\lambda_{1}\right)\left(a(u)X(u) + b(u)\right) + 2\eta \lambda_{1} \overline{v} \right] \\
-\sigma X(u)
\end{matrix}
\right) \, \mathbb{P}\text{-a.s.}, \, u\in [t,T].
\end{equation*}
\end{corollary}

\begin{proof}
See Appendix \ref{app:proofs}, p. \pageref{app:Z-star}.
\end{proof}

\section{Optimal liquidation strategies} \label{sect:opt-liquidation-strategies}

The above solution provides a number of interesting insights on the optimal liquidation policy in relation to the exogenous parameters that define the economic setting. As a general observation it is important to remark that the selection of a dynamic risk measure and the inclusion of a running penalty differentiate substantially the optimal liquidation strategies found in this analysis with respect to those computed considering static measures of risk. In order to best understand these differences, it is useful to restate the optimal control policy in a economically sound way. This is done in the following Proposition.

\begin{proposition} \label{prop:trading-vel}
The optimal trading policy \eqref{eq:v-star} can be specified as follows:
\begin{equation} \label{eq:trading-vel}
v^{\star}(t )-\overline{v}^{\ell}=-\frac{a(t)}{2\left( \eta +\lambda_{1}\right)} \left(X^{\star}(t) -\ell \left( T-t\right) \right), \quad t \in [0,T],
\end{equation}
where:
\begin{equation*}
\overline{v}^{\ell }:=\frac{\lambda_{1}}{\eta +\lambda_{1}} \overline{v},
\end{equation*}
the coefficient $a(\cdot)$ is specified in \eqref{eq:ODEs-solution}, that we recall for convenience:
\begin{equation*}
\begin{array}{l}
a(t)= -\gamma \sqrt{\kappa\lambda_{2}} \frac{2\beta \sqrt{\kappa\lambda_{2}} \sinh \left[ \frac{\gamma \sqrt{\kappa \lambda_{2}}}{2\left(\eta +\lambda_{1}\right)} (T-t)\right] + \left(2\beta -\gamma\right) \cosh \left[ \frac{\gamma \sqrt{\kappa \lambda_{2}}}{2\left(\eta +\lambda_{1}\right)} (T-t)\right]}{\left[2\beta \left(1 -\kappa\lambda_{2} \right) -\gamma\right] \sinh\left[ \frac{\gamma \sqrt{\kappa \lambda_{2}}}{2\left(\eta +\lambda_{1}\right)} (T-t)\right] +\gamma \sqrt{\kappa\lambda_{2}} \cosh \left[ \frac{\gamma \sqrt{\kappa \lambda_{2}}}{2\left(\eta +\lambda_{1}\right)} (T-t)\right]}, \quad t \in [0,T],
\end{array}
\end{equation*}
and the function $\ell \colon [0,T] \rightarrow \mathbb{R}$ reads as:
\begin{equation*}
\ell \left( T-t\right) :=\frac{2\lambda_{1}\overline{v}}{\gamma \sqrt{\kappa\lambda_{2}}}\frac{2\beta \left( \cosh \left[ \frac{\gamma \sqrt{\kappa \lambda_{2}}}{2\left(\eta +\lambda_{1}\right)} (T-t)\right] -1\right) +\frac{(2\beta -\gamma)}{\sqrt{\kappa\lambda_{2}}}\sinh \left[ \frac{\gamma \sqrt{\kappa \lambda_{2}}}{2\left(\eta +\lambda_{1}\right)} (T-t)\right]}{2\beta \sinh \left[ \frac{\gamma \sqrt{\kappa \lambda_{2}}}{2\left(\eta +\lambda_{1}\right)} (T-t)\right] +\frac{(2\beta -\gamma )}{\sqrt{\kappa\lambda_{2}}}\cosh \left[ \frac{\gamma \sqrt{\kappa \lambda_{2}}}{2\left(\eta +\lambda_{1}\right)} (T-t)\right]}.
\end{equation*}
\end{proposition}

\begin{proof}
See Appendix \ref{app:proofs}, p. \pageref{app:trading-vel}.
\end{proof}

In general, the agent will set a liquidation speed deviating from the target one, $\overline{v}^{\ell }$, by an amount proportional to the deviation of the position with respect to a scheduled liquidation program which is defined by the function $\ell \left( T-\cdot\right)$. The mean reversion rate is proportional to $-a(\cdot)$, i.e., the opposite of $a(\cdot)$, which is positive under the sufficient assumptions for the value function to be concave. Indeed, the coefficient $a(\cdot)$ may be equivalently rewritten as:
\begin{equation*}
\begin{array}{l}
a(t)= -\gamma \sqrt{\kappa\lambda_{2}} \frac{2\beta \sqrt{\kappa\lambda_{2}} \tanh \left[ \frac{\gamma \sqrt{\kappa \lambda_{2}}}{2\left(\eta
+\lambda_{1}\right)} (T-t)\right] + \left(2\beta -\gamma\right)}{(2\beta -\gamma) \left(1 -\kappa\lambda_{2} \right) \tanh \left[\frac{\gamma \sqrt{\kappa \lambda_{2}}}{2\left(\eta +\lambda_{1}\right)} (T-t)\right] +\gamma \sqrt{\kappa\lambda_{2}} \left(1 -\sqrt{\kappa\lambda_{2}} \tanh \left[\frac{\gamma \sqrt{\kappa \lambda_{2}}}{2\left(\eta +\lambda_{1}\right)} (T-t)\right]\right]},
\end{array}
\end{equation*}
which is clearly negative, for any $t \in [0,T]$, when $\beta >\frac{\gamma}{2}$ and $\lambda_{2} < \frac{1}{\kappa}$.

Recalling that the effective drift term of the optimal position $X^{\star}(\cdot)$ is proportional to $-v(\cdot)$, this implies that the resulting optimal policy enforces mean reversion toward a scheduled liquidation program. In order to gain intuition on relation \eqref{eq:trading-vel}, it is worth to consider first of all its limiting expression as the risk aversion parameter $\lambda_{2} \rightarrow 0$. We obtain:
\begin{equation*}
\lim_{\lambda_{2}\rightarrow 0}\ell \left( T-t\right) =\overline{v}^{\ell} \left( T-t\right), \quad \forall t \in [0,T],
\end{equation*}
i.e., the target liquidation program for a risk neutral agent corresponds to a constant speed liquidation program. The target velocity does correspond to the one set by the investment committee $\overline{v}$ reduced by a factor $\frac{\lambda_{1}}{\lambda_{1} +\eta}$ that takes into account the effect of the transitory price impact on trades. As expected, in the same limit the mean reversion rate increases with the cost of the final block trade $\beta$ and decreases with the size of permanent impact $\gamma$. Indeed:
\begin{equation*}
\lim_{\lambda_{2}\rightarrow 0}-a(t)=\frac{\left(\eta +\lambda_{1} \right)\left(\beta -\frac{\gamma}{2}\right)}{\left(\eta +\lambda_{1} \right) +(\beta -\frac{\gamma}{2})(T-t)}, \quad t \in[0,T],
\end{equation*}
which is positive considering the admissibility condition $\beta > \frac{\gamma}{2}$. In the limit as $\lambda_{2} \rightarrow 0$, the tracking error induced by the optimal strategy is determined by the ratio $\frac{\beta -\frac{\gamma }{2}}{\left( \eta +\lambda_{1}\right) }$.

For levels of risk aversion that satisfy the sufficient condition for concavity $\lambda_{2}<\frac{1}{\kappa }$, which is equivalent to $\lambda_{2} < \frac{1}{2m^{2}\left( \eta + \lambda_{1}\right)} $ by \eqref{eq:rho}, the optimal policy is essentially modified as follows: a positive (negative) deviation from the target liquidation policy $\overline{v}$ implies a corresponding increase (decrease) of the trading speed with coefficient $-a\left( \cdot\right)$ divided by $\left(\eta +\lambda_{1}\right)$. Note that for finite levels of risk aversion $\lambda_{2}$, the target liquidation program differs substantially from a linear liquidation program. The policy ceases to be linear and the non-linearity increases with the time to the final liquidation. We can identify two well-defined regimes: the late stage liquidation regime and the early stage liquidation regime.

\subsection{Late stage liquidation regime}

In the late liquidation stage corresponding to the limit as $(T-t)\rightarrow 0$ we may assume:
\begin{equation*}
\tanh \left[ \frac{\gamma \sqrt{\kappa \lambda_{2}}}{2\left(\eta +\lambda_{1}\right)} (T-t)\right] \simeq \frac{\gamma \sqrt{\kappa \lambda_{2}}}{2\left( \eta +\lambda_{1}\right) }(T-t)=\frac{\gamma m\sqrt{\lambda_{2}}}{\sqrt{2\left( \eta +\lambda_{1}\right) }}(T-t).
\end{equation*}
In this case, for any $t\in [0,T]$, we obtain:
\begin{equation*}
\begin{aligned} \lim_{(T -t) \rightarrow 0}\ell \left( T-t\right) &= \frac{\overline{v}^{\ell}(T-t)}{1 +\frac{1}{\gamma^{-1} -\left( 2\beta \right)^{-1}}\lambda_{2}m^{2}(T-t)} =: \ell_{0}, \\
\lim_{(T -t) \rightarrow 0} \left(-a (t)\right) &= 2\left( \eta +\lambda_{1}\right) \frac{2\beta
-\gamma +2\beta \gamma m^{2}\lambda_{2}(T-t)}{\left[ 2\beta -\gamma-2\beta m^{2}\lambda_{2}\left( \eta +\lambda_{1}\right) \right] (T-t)+2\left( \eta
+\lambda_{1}\right) } =: -a_{0}.
\end{aligned}
\end{equation*}
This regime corresponds to the one where, setting $\lambda_{1}=\lambda_{2}=0$, one recovers the Adaptive Value Weighted Average Price (AVWP) strategy found in \cite{Xue}. The contribution induced by $\lambda_{1}>0$ changes the benchmark liquidation policy, whereas the parameter $\lambda_{2}>0$ in this regime operates simply as a renormalization of the reference strategy, progressively reducing the liquidation velocity far from the final block liquidation date while raising the mean reversion rate.

\subsection{Early stage liquidation regime}

In the early liquidation stage corresponding to the limit $(T-t)\rightarrow \infty $ regime we have:
\begin{equation*}
\begin{aligned}
\lim_{(T -t) \rightarrow +\infty}\ell \left( T-t\right) &= \frac{2\lambda_{1}\overline{v}}{\gamma m\sqrt{2\lambda_{2}\left( \eta +\lambda_{1}\right)}}=: \ell_{\infty}, \\
\lim_{(T -t) \rightarrow +\infty} \left(-a (t)\right) &= \frac{\gamma m\sqrt{2\lambda_{2}\left( \eta +\lambda_{1}\right)}}{1 -m\sqrt{2\lambda_{2}\left( \eta +\lambda_{1}\right)}} =: -a_{\infty}.
\end{aligned}
\end{equation*}
This is the regime where the impact of the backward component and of the penalization for deviations of the trading velocity is more evident. Far from the final block liquidation the presence of the risk averse component drives the optimal control policy into a steady policy depending on the size of the permanent impact $\gamma$, but independent from the cost $\beta$ of the final block order. In this regime, the optimal evolution for the process $X^{\star}(\cdot)$ is given by:
\begin{equation*}
dX^{\star}(t) =-\frac{a_{\infty }}{2\left( \eta +\lambda_{1}\right)} \left(\overline{x}_{\infty }-X^{\star}(t)\right) dt+mdB_{1}(t), \quad t \in [0,T],
\end{equation*}
where:
\begin{equation*}
\overline{x}_{\infty } := \frac{-a_{\infty }\ell_{\infty} -\lambda_{1} \overline{v}}{-\frac{a_{\infty }}{2\left( \eta +\lambda_{1}\right)}},
\end{equation*}
i.e.,
\begin{equation*}
\overline{x}_{\infty } = \frac{2}{\gamma}\left( \eta +\lambda_{1}\right) \left( \frac{1}{m\sqrt{2\lambda_{2}\left( \eta +\lambda_{1}\right)}} + 1 \right) \lambda_{1}\overline{v},
\end{equation*}
which corresponds to a standard mean reverting Ornstein-Uhlenbeck process with rate of mean reversion $-\frac{a_{\infty }}{2\left( \eta +\lambda_{1}\right)}$ vanishing in the limit as $\lambda_{2}m\gamma \rightarrow 0$. In other words, if the final block trade is sufficiently far in the future, the investor is simply controlling the quantity of security held with a tracking error that is decreasing with increasing permanent impact $\gamma$, risk aversion $\lambda_{2}$ and volatility of the position $m$.

Notice that in this regime the higher the price impact the higher the optimal reaction of the trader to a deviation from the optimal path. In fact, the presence of the uncertainty of order filling generates a cost that is increasing with increasing price impact. In the absence of the backward component, i.e., $\lambda_{2}=0$, the trader would not be penalized for this term. In other words, it is possible to interpret the penalization component induced by the backward part, as a term that penalizes those strategies inducing high tracking errors in the liquidation strategy relative to the benchmark set by the investment committee.

In summary, it is interesting to remark that under the conditions $\lambda_{1}>0$ and $\lambda_{2}>0$, the liquidation policy smoothly interpolates between a stationary investment committee policy when $(T-t) \rightarrow +\infty$, and a liquidation policy with constant rate of liquidation set by the investment committee with a mean reversion rate that increases with increasing risk aversion $\lambda_{2}$.

In order to better illustrate the impact of the backward component and of the risk aversion parameter $\lambda_{2}$ on the liquidation strategy, we provide a graphical illustration of the target liquidation schedule and of the mean reversion rate for different levels of $\lambda_{2}$ as a function of time.

\begin{figure}[h]
\centering
\includegraphics[scale=1.0]{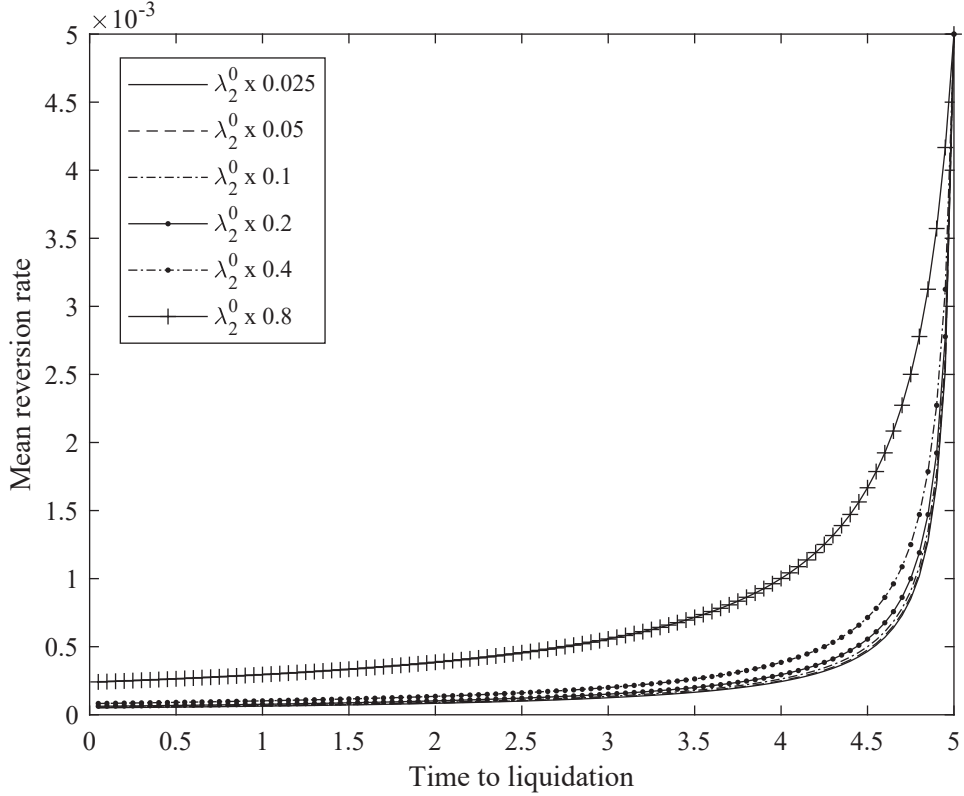}
\caption{Mean reversion rate $-a\left( \cdot\right) $ as a function of time. A higher risk aversion implies a faster increase of the mean reversion rate. Traders with greater risk aversion agree to pay a higher cost in order to reduce deviations with respect to the scheduled liquidation program. We set $T=5$ as maximum allowed time limit.} \label{fg:1}
\end{figure}

In Figure \ref{fg:1} we perform a numerical illustration using the parameters similar to the one used in \cite{Xue} which are closely related to those discussed in \cite{Almgren-Chriss}:
\begin{gather*}
\gamma =2.5\times 10\symbol{94}(-7),\qquad \eta =25\times 10\symbol{94}(-6), \qquad m=0.2\times 10\symbol{94}(7), \qquad \beta =100\times \eta , \\
\lambda_{1}=0.0001, \qquad \lambda_{2}^{0}:=2m^{2}\left( \lambda_{1} +\eta\right) = 10\symbol{94}\left( -6\right).
\end{gather*}

As we should expect, a higher risk aversion implies that the agent will try to reduce the tracking error implementing a policy where mean reversion raises earlier and faster tightening the policy reaction to deviations from the scheduled liquidation program. As a matter of fact, we see that the mean reversion is increasing with time and with parameter $\lambda_{2}$.

A critical innovation of the present approach compared to previous liquidation models is that the parameter $\lambda_{2}$ determines directly the characteristic duration of the trade $\frac{\gamma \sqrt{\kappa\lambda_{2}}}{2\left( \eta +\lambda_{1}\right)}$ jointly with the size of the permanent impact $\gamma$, the volatility of the order fills $m$, and the quantity $(\eta +\lambda_{1})$.

While in the conventional framework of \cite{Almgren-Chriss} the characteristic ``trade time'' is set by a specific parameter, in this case this characteristic time is a function of a number of parameters, including trader's subjective risk aversion $\lambda_{2}$ and the volatility coefficient $m$ that quantifies uncertainty of order fills.

\begin{figure}[h]
\centering
\includegraphics[scale=1.0]{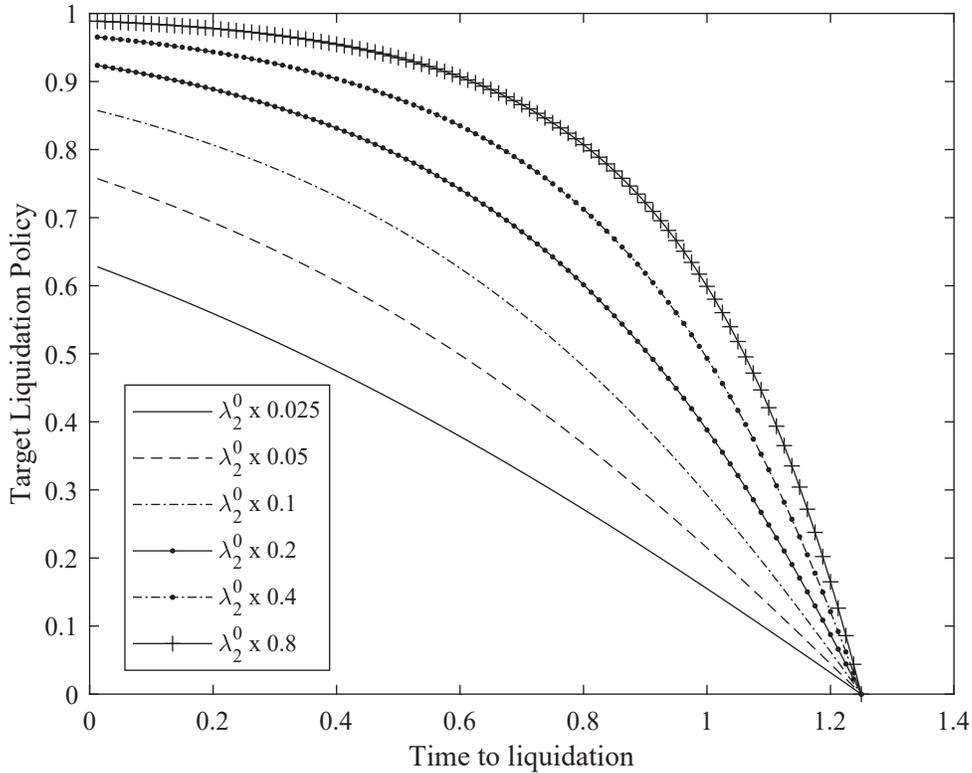}
\caption{The scheduled liquidation program for traders with different degree of risk aversion. Time variable is scaled by a factor $T_{0} = \frac{\sqrt{\kappa \lambda_{2}^{0}}}{2\left(\eta + \lambda_{1}\right)}$ in order to express it in units proportional to the characteristic trading time.} \label{fg:2}
\end{figure}

In Figure \ref{fg:2}, to make comparable the target liquidation paths for different levels of $\lambda_{2}$, we choose to set the target $\overline{v}$ as a value that normalizes this asymptotic early stage position to a reference value equal to $1$, corresponding to a $100\%$ notional amount to be liquidated. In addition, we consider a longer liquidation horizon to magnify the relationship between the level of risk aversion and the concavity of the scheduled liquidation program that converges to a linear liquidation program as the final block trade date is approached. The dynamic nature of the risk adjustment raises the importance of the deviation from the scheduled target, while reducing the relative importance of the cost paid by the trader in the final block trade. This implies a raise of concavity of the liquidation path, which signals the transition between early and late liquidation regimes.

Note that a joint interpretation of the evidence coming from Figures \ref{fg:1} and \ref{fg:2} indicates that the minimization of the volatility of the backward component induced by the dynamic risk measure determines a policy and a liquidation schedule that are progressively and non-linearly tightening as the final liquidation is approached. In light of this consideration and the above results, an interesting argument that future research will have to clarify is the economic interpretation of the characteristic ``trading time'' resulting from the interaction between the uncertainty of order fills, the risk aversion parameter of the dynamic risk measures, and characteristic half life of the trade.

%
%

\section*{Acknowledgement}

We wish to thank two anonymous referees and the editors for careful scrutiny and helpful suggestions. We thanks Claudio Tebaldi for valuable discussions and for carefully reading the draft of this paper. Holger Kraft, Athena Picarelli and Emanuela Rosazza-Gianin deserve special mention for their useful comments. Finally, we are grateful to all participants to seminars and conferences where the work was presented. The usual disclaimer applies.

%
%

%
%

\appendix

\section{Some technical details on the profit and loss function} \label{app:P&L}

We provide to give more technical detail about the profit and loss function.

\subsection{Decomposition of the P\&L functional}
\label{app:decomposition-P&L}

It is important to observe that a properly defined P\&L must admit a decomposition into two contributions: one can be regarded as a modified self-financing strategy proposed in \cite{Carmona-Webster} and  the other corresponds to slippage. The formula must recover the classical self-financing condition in the absence of trading frictions\footnote{We thank an anonymous referee for pointing this out.}.

The P\&L defined in \eqref{eq:P&L} that we recall for convenience:
\begin{equation*}
\Pi^{0}(t) := X(t)\left( S(t)-S(0)\right) +\int_{0}^{t} (S(0)-\widetilde{S}(u)) dX(u),
\end{equation*}
can be decomposed in a self-financing strategy contribution and a slippage component. Moreover, it is consistent with the self-financing condition introduced by \cite[p.~731]{Carmona-Webster}. They generalize the usual self-financing relationships of frictionless markets to make it compatible with markets with frictions, including the presence of the uncertainty in the order fills as defined in our model.

In the following, we show how to decompose the above P\&L formula by adding and subtracting the terms $\int_{0}^{t} S(u) dX(u)$ to its right-hand side. We obtain:
\begin{equation*}
\begin{aligned}
\Pi ^{0}(t) &= X(t) (S(t) - S(0)) + \int_{0}^{t} (S(0) - \widetilde{S}(u)) dX(u) = \\
&= X(t)S(t) - X(0)S(0) -\int_{0}^{t}\widetilde{S}(u)dX(u) + \int_{0}^{t} S(u) dX(u) - \int_{0}^{t} S(u) dX(u),
\end{aligned}
\end{equation*}
that can be decomposed as:
\begin{equation*}
\Pi ^{0}(t) = \underset{\text{modified self-financing strategy}}{\underbrace{X(t) S(t) - X(0) S(0) - \int_{0}^{t} S(u) dX(u)}} +\underset{\text{slippage}}{\underbrace{\int_{0}^{t} (S(u) - \widetilde{S}(u)) dX(u)}}.
\end{equation*}
In fact, integration by parts implies:
\begin{equation*}
X(t) S(t)- X(0)S(0) -\int_{0}^{t}S(u) dX(u) = \int_{0}^{t} X(u)dS(u)+ \int_{0}^{t}\langle dS, dX\rangle (u)
\end{equation*}
that is the value accrued by trading on $S(u)$ using a self-financing strategy $X(u)$, for any $u \in [0,t]$, according to the definition extended by \cite{Carmona-Webster} to take into account trading frictions. Correspondingly, the amount:
\begin{equation*}
\int_{0}^{t}(S(u) - \widetilde{S}(u)) dX(u)
\end{equation*}
can be interpreted as a slippage component since it properly vanishes as soon as the price impact is set to zero, i.e., $\widetilde{S}(u) = S(u)$, for any $u\in [0,t]$.

\subsection{Computation of the P\&L functional}
\label{app:computation-P&L}

We have:
\begin{equation} \label{eq:P&L-intermidiate}
\begin{aligned}
\Pi^{0}(t) :=& X(t) (S(t) - S(0)) +\int_{0}^{t} (S(0) - \widetilde{S}(u)) dX(u) =\\
=& X(t) (S(t) - S(0)) -\int_{0}^{t} (S(u) - S(0)) dX(u) +\eta \int_{0}^{t} v(u) dX(u) =\\
=& \int_{0}^{t} X(u)dS(u) + \int_{0}^{t} \langle dS, dX \rangle (u) -\eta \int_{0}^{t} v^{2}(u) du + \\
&+ \eta m \int_{0}^{t} v(u)dB_{1}(u) = \\
=&  \gamma m^{2} t + \int_{0}^{t} X(u)dS(u) -\eta \int_{0}^{t} v^{2}(u) du + \eta m\int_{0}^{t} v(u)dB_{1}(u) =\\
=& \gamma m^{2} t + \gamma\int_{0}^{t} X(u)dX(u) -\eta \int_{0}^{t} v^{2}(u)du + \eta m\int_{0}^{t} v(u)dB_{1}(u) + \\
& +\sigma \int_{0}^{t} X(u) dB_{2}(u).
\end{aligned}
\end{equation}
Since:
\begin{equation*}
dX^{2}(u) = 2X(u)dX(u) + m^{2} du \quad \Longleftrightarrow \quad X(u)dX(u) = \frac{1}{2} \left(dX^{2}(u) - m^{2}du\right),
\end{equation*}
then:
\begin{equation*}
\begin{aligned}
\Pi^{0}(t) =& \gamma m^{2} t + \gamma\int_{0}^{t} X(u)dX(u) -\eta \int_{0}^{t} v^{2}(u)du + \eta m\int_{0}^{t} v(u)dB_{1}(u) + \\
& +\sigma \int_{0}^{t} X(u) dB_{2}(u) =\\
=& \gamma m^{2} t + \frac{\gamma}{2} \int_{0}^{t} dX^{2}(u) -\frac{\gamma}{2}m^{2}\int_{0}^{t}du -\eta \int_{0}^{t} v^{2}(u)du + \eta m\int_{0}^{t} v(u) dB_{1}(u) +\\
&+ \sigma \int_{0}^{t} X(u) dB_{2}(u) =\\
=& \frac{\gamma}{2} \left(X^{2}(t) -x_{0}^{2}\right) +\frac{\gamma}{2} m^{2}t -\eta \int_{0}^{t} v^{2}(u)du + \eta m\int_{0}^{t} v(u) dB_{1}(u) +\\
&+ \sigma \int_{0}^{t} X(u) dB_{2}(u) .
\end{aligned}
\end{equation*}
Equivalently, taking into account the last equality of \eqref{eq:P&L-intermidiate} we obtain:
\begin{equation*}
\begin{aligned}
\Pi^{0}(t) =& \gamma m^{2} t + \gamma\int_{0}^{t} X(u)dX(u) +\sigma \int_{0}^{t} X(u) dB_{2}(u) -\eta \int_{0}^{t} v^{2}(u)du + \\
&+ \eta m\int_{0}^{t} v(u)dB_{1}(u) =\\
=& \gamma m^{2} t -\gamma\int_{0}^{t} v(u)X(u)du +\gamma m \int_{0}^{t} X(u)dB_{1}(u) +\sigma \int_{0}^{t} X(u) dB_{2}(u) +\\
&- \eta \int_{0}^{t} v^{2}(u)du + \eta m\int_{0}^{t} v(u)dB_{1}(u) =\\
=& \gamma m^{2} t -\int_{0}^{t} \left(\gamma v(u)X(u) + \eta v^{2}(u)\right)du + m\int_{0}^{t} \left(\gamma X(u)+ \eta v(u)\right) dB_{1}(u) +\\
&+\sigma \int_{0}^{t} X(u) dB_{2}(u).
\end{aligned}
\end{equation*}

\section{Technical proofs} \label{app:proofs}

Here we provide the technical proofs.

\subsection*{Proof of Proposition \ref{prop:erm}} \label{app:erm}

It follows straightforward from the result showed in \cite[Proposition~3.12, p.~123]{Barrieu-ElKaroui} and the comparison theorem presented in \cite[Theorem~5, p.~554]{Briand-Hu2}. \qed

\bigskip

\subsection*{Semigroup property of Remark \ref{rmk:sg-property}} \label{app:sg-property}

We have:
\begin{equation*}
\begin{aligned}
&\mathcal{R}\left(t,\mathcal{R}(\tau,\xi(T))\right) = \\
&=\mathcal{R}\left(\tau,\xi(T) \right) +\int_{t}^{\tau} g\left(Z_{1}(u),Z_{2}(u)\right)du -\int_{t}^{\tau} Z_{1}(u)dB_{1}(u) -\int_{t}^{\tau} Z_{2}(u) dB_{2}(u) = \\
&= -\xi(T) +\int_{\tau}^{T} g\left(Z_{1}(u),Z_{2}(u)\right)du -\int_{\tau}^{T} Z_{1}(u) dB_{1}(u) -\int_{\tau}^{T} Z_{2}(u) dB_{2}(u) +\\
&\quad +\int_{t}^{\tau} g\left(Z_{1}(u),Z_{2}(u)\right)du -\int_{t}^{\tau}Z_{1}(u) dB_{1}(u) -\int_{t}^{\tau} Z_{2}(u) dB_{2}(u) = \\
& = -\xi(T) +\int_{t}^{T} g\left(Z_{1}(u), Z_{2}(u)\right)du -\int_{t}^{T} Z_{1}(u) dB_{1}(u) -\int_{t}^{T} Z_{2}(u)dB_{2}(u) =\\[-3pt]
&=\mathcal{R}\left(t,\xi(T)\right).
\end{aligned} \qed
\end{equation*}

\bigskip

\subsection*{Proof of Lemma \ref{lemma:HJB-solution}} \label{app:HJB-solution}

The claim follows by direct computations and observing that:
\begin{equation*}
\begin{array}{l}
a(t) -\gamma= -\gamma \frac{(2\beta -\gamma) \sinh \left[\frac{\gamma \sqrt{\kappa \lambda_{2}}}{2\left(\eta +\lambda_{1}\right)} (T-t)\right] +2\beta \sqrt{\kappa\lambda_{2}}\cosh \left[\frac{\gamma \sqrt{\kappa \lambda_{2}}}{2\left(\eta +\lambda_{1}\right)} (T-t)\right]}{ \left[2\beta \left(1 -\kappa\lambda_{2}\right) -\gamma \right] \sinh \left[\frac{\gamma \sqrt{\kappa \lambda_{2}}}{2\left(\eta +\lambda_{1}\right)} (T-t)\right] + \gamma
\sqrt{\kappa\lambda_{2}}\cosh \left[\frac{\gamma \sqrt{\kappa \lambda_{2}}}{2\left(\eta +\lambda_{1}\right)} (T-t)\right]}
\end{array}
\end{equation*}
or, equivalently,
\begin{equation*}
\begin{array}{l}
a(t) -\gamma= -\gamma \frac{(2\beta -\gamma) \tanh \left[\frac{\gamma \sqrt{\kappa \lambda_{2}}}{2\left(\eta +\lambda_{1}\right)} (T-t)\right] + 2\beta
\sqrt{\kappa\lambda_{2}}}{(2\beta -\gamma) \left(1 -\kappa\lambda_{2} \right) \tanh \left[\frac{\gamma \sqrt{\kappa \lambda_{2}}}{2\left(\eta +\lambda_{1}\right)} (T-t)\right] +\gamma \sqrt{\kappa\lambda_{2}} \left(1 -\sqrt{\kappa\lambda_{2}} \tanh \left[\frac{\gamma \sqrt{\kappa \lambda_{2}}}{2\left(\eta +\lambda_{1}\right)} (T-t)\right]\right]},
\end{array}
\end{equation*}
is clearly negative if $\beta > \dfrac{\gamma}{2}$ and $\lambda_{2} < \dfrac{1}{\kappa}$, since $\frac{\gamma \sqrt{\kappa \lambda_{2}}}{2\left(\eta +\lambda_{1}\right)} (T-t) \geq 0$, for any $t \in [0, T]$.

Regarding the solution to the system of Riccati ODEs, we point out that the computation for the functions $a(\cdot)$ and $c(\cdot)$ comes straightforward from solving the associated ODEs, while $b(\cdot)$ is recovered by the explicit computation of the solution to the ODE derived for the function $\ell \colon [0,T] \rightarrow \mathbb{R}$, which is defined as:
\begin{equation*}
\ell\left(T-t\right) := -\frac{b(t)}{a(t)}, \quad t \in[0,T].
\end{equation*}
Indeed, the flow generated by the ODEs for $a(\cdot)$ and $b(\cdot)$ induces a linear ODE for $\ell(T-\cdot)$ that can be solved by variation of constants.\footnote{Details of the (long) computation are available upon request.} \qed

\bigskip

\subsection*{Proof of Theorem \ref{th:vt}} \label{app:vt}

We know that the function $w$ given in \eqref{eq:ansatz} satisfies the generalized HJB equation \eqref{eq:HJB-scalar}-\eqref{eq:HJB-tc} for any $x \in \mathbb{R}$ and want to prove that this solution in this case is the unique value function for the optimal control problem. Let us consider $x \in \mathbb{R}$ and $v\left( \cdot \right) \in \mathcal{V}_{ad}\left[t,T\right]$ with the associated state trajectory $X(\cdot) := X(\cdot; t,x,v)$. Apply the Dynkin formula to the functions $(t,x) \mapsto \frac{1}{2}\left( a(t)-\gamma \right) x^{2}$ and $(t,x) \mapsto b(t)x$ with the process $X(\cdot)$, respectively. For any $u\in [ t,T]$, we obtain:
\begin{multline*}
\mathbb{E}\left[ \int_{t}^{T}d\left( \frac{1}{2}\left[ a(u)-\gamma \right] X^{2}(u)\right) \right] = \\
=\mathbb{E}\left[ \int_{t}^{T}\left\{ \frac{1}{2}\dot{a}(u)X^{2}(u) -\left[a(u) -\gamma \right] X(u)v(u) +\frac{1}{2}m^{2}a(u)\right\} du\right] ,
\end{multline*}
then:
\begin{multline} \label{eq:final-condition}
\mathbb{E}\left[ -\beta X^{2}(T)\right] = \\
=\frac{1}{2}\left( a(t)-\gamma \right) x^{2}+\mathbb{E}\left[ \int_{t}^{T}\left\{ \frac{1}{2}\left[ \lambda_{2}m^{2}-\frac{1}{2\left( \eta
+\lambda_{1}\right) }\right] a^{2}(u)X^{2}(u)+\right. \right. \\
-\lambda_{2}m^{2}\gamma a(u)X^{2}(u)+\frac{1}{2}\lambda_{2}m^{2}\gamma^{2} X^{2}(u) -a(u)X(u)v(u) +\gamma X(u)v(u) + \\
+\frac{1}{2}m^{2}a(u)-\frac{1}{2}m^{2}\gamma \bigg\}du\bigg]
\end{multline}
and
\begin{equation*}
\mathbb{E}\left[ \int_{t}^{T}d\left( b(u)X(u)\right) \right] =\mathbb{E} \left[ \int_{t}^{T}\left( \dot{b}(u)X(u)-b(u)v(u)\right) du\right] ,
\end{equation*}
i.e.,
\begin{multline} \label{eq:w-linear-term}
-b(t)x = \mathbb{E}\left[ \int_{t}^{T}\left\{ \left[ \lambda_{2}m^{2} -\frac{1}{2\left( \eta +\lambda_{1}\right) }\right] a(u)b(u)X(u) +\frac{\lambda_{1}
\overline{v}}{\eta +\lambda_{1}}a(u)X(u) +\right. \right. \\
-\lambda_{2}m^{2}\gamma X(u) -b(u)v(u)\bigg\}du\bigg].
\end{multline}
Moreover:
\begin{multline*}
\mathbb{E}\left[ \int_{t}^{T}d\left( c(u)\right) \right] = \mathbb{E}\left[ \int_{t}^{T}\left\{ \dfrac{1}{2}\left[ \lambda_{2}m^{2}-\dfrac{1}{2\left( \eta +\lambda_{1}\right) }\right] b^{2}(t) +\dfrac{\lambda_{1} \overline{v}}{\eta +\lambda_{1}}b(t) +\right. \right. \\
\left. \left. -\dfrac{1}{2}m^{2}a(t) -\frac{1}{2}m^{2}\gamma +\lambda_{1} \overline{v}^{2} -\dfrac{\lambda_{1}^{2}\overline{v}^{2}}{\eta +\lambda_{1}}\right\} du\right] ,
\end{multline*}
i.e.,
\begin{multline} \label{eq:w-constant-term}
-c(t) = \mathbb{E}\left[ \int_{t}^{T}\left\{ \dfrac{1}{2}\left[ \lambda_{2} m^{2} -\dfrac{1}{2\left( \eta +\lambda_{1}\right)} \right] b^{2}(t) +\dfrac{\lambda_{1} \overline{v}}{\eta +\lambda_{1}}b(t) -\dfrac{1}{2} m^{2}a(t) +\right. \right. \\
\left. \left. -\frac{1}{2}m^{2}\gamma +\lambda_{1} \overline{v}^{2} -\dfrac{\lambda_{1}^{2} \overline{v}^{2}}{\eta +\lambda_{1}}\right\} du \right]
\end{multline}

Recalling \eqref{eq:functional}, the objective functional can be recast as:
\begin{equation*}
\begin{aligned} J(t,x;v(\cdot)) &= Y(t;t,x;v(\cdot)) = \mathbb{E} \left[Y(t;t,x;v(\cdot))\right] = \\ &= \mathbb{E} \left[-\beta X^{2}(T) -\int_{t}^{T} \left(\gamma X(u) -2\lambda_{1}\overline{v}\right) v(u)du -\left(\lambda_{1} +\eta\right) \int_{t}^{T} v^{2}(u)du + \right. \\
&\quad \left. -\dfrac{\lambda_{2}}{2} \int_{t}^{T} \left(Z_{1}^2(u)+Z_{2}^2(u)\right) du\right] +\left(\gamma m^{2} -\lambda_{1} \overline{v}^{2}\right) \left(T-t\right),
\end{aligned}
\end{equation*}
Thus, by substituting \eqref{eq:final-condition} into the above we have the following:
\begin{eqnarray*}
J(t,x;v(\cdot )) &=&\frac{1}{2}\left( a(t)-\gamma \right) x^{2}+\mathbb{E} \left[ \int_{t}^{T}\bigg\{-\left( \lambda_{1}+\eta \right) v^{2}(u)+\right. \\
&&-\left( a(u)X(u)-2\lambda_{1}\overline{v}\right) v(u)-\dfrac{\lambda_{2}}{2} \left(\widetilde{Z}_{1}^{2}(u) +\widetilde{Z}_{2}^{2}(u)\right) + \\
&&+\frac{1}{2}\left[ \lambda_{2}m^{2}-\frac{1}{2\left( \eta +\lambda_{1}\right) }\right] a^{2}(u)X^{2}(u) -\lambda_{2}m^{2}\gamma a(u)X^{2}(u)+ \\
&&+\frac{1}{2}\lambda_{2}m^{2} \gamma^{2}X^{2}(u) +\frac{1}{2}m^{2}a(u) +\frac{1}{2}m^{2}\gamma -\lambda_{1}\overline{v}^{2}\bigg\}du \bigg],
\end{eqnarray*}
from which, applying \eqref{eq:w-linear-term}, recalling \eqref{eq:w-constant-term}, and reorganizing all the terms, we obtain:
\begin{equation} \label{eq:w-whole-terms}
\begin{aligned}
&J(t,x;v(\cdot)) =\frac{1}{2}\left(a(t) -\gamma\right)x^{2} +b(t)x +c(t) + \\ &+ \mathbb{E}\left[\int_{t}^{T} \left\{-\frac{1}{4\left(\eta +\lambda_{1}\right)} \left[a(u)X(u) +b(u) -2\lambda_{1} \overline{v} + 2\left(\eta +\lambda_{1}\right)v(u)\right]^{2} +\right.\right. \\
&\left.\left. +\frac{\lambda_{2}}{2} \left\{m^{2}\left[\left(a(u) -\gamma \right) X(u) +b(u)\right]^{2} -\left(\widetilde{Z}_{1}^{2}(u) +\widetilde{Z}_{2}^2(u)\right) \right\} \right\} du\right].
\end{aligned}
\end{equation}

For any fixed time $u\in [ t,T]$, let us define the following function:
\begin{multline} \label{eq:L-function}
L^{u}(v,\lambda_{2}):=-\frac{1}{4\left( \eta +\lambda_{1}\right) }\left[ a(u)X(u)+b(u)-2\lambda_{1}\overline{v}+2\left( \eta +\lambda_{1}\right) v \right]^{2}+ \\
+ \frac{\lambda_{2}}{2}\left\{m^{2}\left[ \left( a(u)-\gamma \right) X(u)+b(u)\right]^{2} - (\widetilde{Z}_{1}^{2} +\widetilde{Z}_{2}^{2}) \right\} ,
\end{multline}
which is by definition equal to the argument of the integral in the right side hand of \eqref{eq:w-whole-terms} at given time $u\in [t,T]$. It has the interpretation of the Lagrangian function related to the following constrained optimization problem:
\begin{gather*}
\sup_{v\in \mathbb{R}}\left\{ -\frac{1}{4\left( \eta +\lambda_{1}\right) } \left[ a(u)X(u) +b(u) -2\lambda_{1} \overline{v}+2\left( \eta +\lambda_{1}\right) v\right]^{2}\right\} \\
\text{subject to} \; (\widetilde{Z}_{1}^{2} +\widetilde{Z}_{2}^{2}) - m^{2}\left[ \left( a(u)-\gamma \right) X(u)+b(u)\right]^{2} \leq 0,
\end{gather*}
that corresponds to the maximization of the contribution driven by the P\&L component of the objective functional for a fixed maximum threshold of risk. Note that the above is a static constrained optimization problem, where the risk aversion parameter $\lambda_{2}$ plays the role of the Lagrangian parameter, $v\in \mathbb{R}$ is the control variable, and the two-dimensional vector $(\widetilde{Z}_{1},\widetilde{Z}_{2})^{\top} \in \mathbb{R}^{2}$ appears as an independent variable.

Let the optimal control variable corresponding to $\lambda_{2}$ be denoted by $v^{\star }(\lambda_{2})$ and $(Z_{1}^{\star }(\lambda_{2}), Z_{2}^{\star}(\lambda_{2}))^{\top}$ be the corresponding risk which satisfies the constraint with equality, the first order conditions for the Lagrangian
function $L^{u}$ defined in \eqref{eq:L-function} read as:
\begin{equation*}
\begin{cases}
\dfrac{\partial L^{u}(v,\lambda_{2})}{\partial v}=-\dfrac{1}{2\left( \eta +\lambda_{1}\right) }\left[ a(u)X(u)+b(u)-2\lambda_{1}\overline{v}+2\left( \eta +\lambda_{1}\right) v\right] =0, \\[9pt]
\dfrac{\partial L^{u}(v,\lambda_{2})}{\partial \lambda_{2}} = \widetilde{Z}_{1}^{2}+\widetilde{Z}_{2}^{2}-m^{2}\left[ \left( a(u)-\gamma \right) X(u)+b(u)\right]^{2}=0,
\end{cases}
\end{equation*}
which imply:
\begin{eqnarray*}
v^{\star }(\lambda_{2}) &=&-\frac{a(u)X(u)+b(u)-2\lambda_{1} \overline{v}}{2\left( \eta +\lambda_{1}\right) }, \\
(\widetilde{Z}_{1}^{\star }(\lambda_{2}))^{2}+(\widetilde{Z}_{2}^{\star}(\lambda_{2}))^{2} &=&m^{2}\left[ \left( a(u)-\gamma \right) X(u)+b(u)\right]^{2}.
\end{eqnarray*}
Since the target function is concave in $v$, we observe that the above first order conditions are necessary and sufficient to select the (constrained) maximum. Moreover, arguing as in \cite[Section~5.2, p.~2972]{Oksendal-Sulem}, we show that:
\begin{equation*}
\!\!\!\sup_{\substack{ v\in \mathbb{R}  \\ \text{s.t.}\;\widetilde{Z}_{1}^{2} +\widetilde{Z}_{2}^{2}-m^{2}\left[ \left( a(u)-\gamma \right) X(u) +b(u)\right]^{2}\leq 0}}\!\!\!\left\{ -\tfrac{1}{4\left( \eta +\lambda_{1}\right)} \left[ a(u)X(u) +b(u) -2\lambda_{1}\overline{v} +2\left( \eta +\lambda_{1}\right) v\right]^{2}\right\}
\end{equation*}
\begin{multline*}
\leq \sup_{v\in \mathbb{R}}\left\{ -\tfrac{1}{4\left( \eta +\lambda_{1}\right) }\left[ a(u)X(u) +b(u) -2\lambda_{1} \overline{v} +2\left( \eta +\lambda_{1}\right) v\right]^{2}\right\} + \\
+\tfrac{\lambda_{2}}{2}\left\{ m^{2}\left[ \left( a(u)-\gamma \right) X(u) +b(u)\right]^{2} -(\widetilde{Z}_{1}^{2} +\widetilde{Z}_{2}^{2}) \right\}
\end{multline*}
\begin{multline*}
=\left\{ -\tfrac{1}{4\left( \eta +\lambda_{1}\right)} \left[a(u)X(u) +b(u) -2\lambda_{1}\overline{v}+2\left( \eta +\lambda_{1}\right) v^{\star }(\lambda_{2})\right]^{2}\right\} + \\
+\tfrac{\lambda_{2}}{2}\left\{ m^{2}\left[ \left( a(u)-\gamma \right) X(u)+b(u)\right]^{2}-\left[ (\widetilde{Z}_{1}^{\star} (\lambda_{2}))^{2} +(\widetilde{Z}_{2}^{\star }(\lambda_{2}))^{2}\right] \right\}
\end{multline*}
\begin{multline*}
=\left\{ -\tfrac{1}{4\left( \eta +\lambda_{1}\right)} \left[ a(u)X(u) +b(u) -2\lambda_{1}\overline{v} +2\left( \eta +\lambda_{1}\right) v^{\star}
(\lambda_{2})\right]^{2} \right\} \\[12pt]
\leq \!\!\!\sup_{\substack{ v\in \mathbb{R}  \\
\text{s.t. } \widetilde{Z}_{1}^{2} +\widetilde{Z}_{2}^{2} -m^{2}\left[ \left( a(u) -\gamma \right) X(u) +b(u)\right]^{2}\leq 0}}\!\!\!\left\{ \!-\tfrac{1}{4\left( \eta +\lambda_{1}\right) }\left[ a(u)X(u) \!+\! b(u) \!-\! 2\lambda_{1}\overline{v} \!+\! 2\left(\eta \!+\!\lambda_{1}\right) v\right]^{2}\right\} .
\end{multline*}
Hence, the maximization of the target function subject to a maximum risk constraint is equivalent to the unconstrained maximization of $L^{u}(v,\lambda_{2})$ for each given $\lambda_{2}>0$.

Now, recall that starting from $x \in \mathbb{R}$ at time $t \in[0,T] $, for each control $v(\cdot) \in \mathcal{V}_{\text{ad}}[t,T]$, there is a unique choice of the process $(\widetilde{Z}^{\star}_{1}, \widetilde{Z}^{\star}_{2})^{\top} (\cdot;t,x;v(\cdot))$ that makes the solution $Y^{\star}(\cdot;t,x;v(\cdot))$ to the backward part of the controlled state equation \eqref{eq:qgFBSDE} adapted, when $X^{\star}(\cdot) := X^{\star}(\cdot;t,x;v(\cdot))$ is the solution to the forward part of \eqref{eq:qgFBSDE}. By \cite[Section~5]{Briand-Hu2} we observe that the backward component in \eqref{eq:qgFBSDE} admits a Feynman-Kac representation; thus, denoting by $w^{v}:[0,T] \times \mathbb{R}\times\mathbb{R} \rightarrow \mathbb{R}$ the solution to a proper semi-linear parabolic PDE, we notice that $w^{v}(t,x) := Y^{\star}(t;t,x,v)$ is a deterministic function, and by the Markov property of the diffusion process we have:
\begin{equation*}
Y^{\star}(t)=w^{v}(t,X^{\star}(t)), \qquad \left(
\begin{matrix}
\widetilde{Z}_{1}^{v} \\
\widetilde{Z}_{2}^{v}
\end{matrix}
\right)(t) = -\Sigma^{\top} \operatorname{D} w^{v}(t,X(t)) , \qquad t\in [ 0,T],
\end{equation*}
where $\Sigma$ is defined in \eqref{eq:vol-matrix}.

Remarkably, at every fixed time $u \in [t,T]$, it is straightforward to verify that $(\widetilde{Z}_{1}^{\star}(\lambda_{2}), \widetilde{Z}_{2}^{\star}(\lambda_{2}) )^{\top} = (\widetilde{Z}_{1}^{v^{\star}}, \widetilde{Z}_{2}^{v^{\star}})^{\top} (u)$, i.e., the first order conditions for the Lagrangian problem are verified by $w^{v^{\star}}(t,x)$ and
\begin{equation*}
w^{v^{\star}}(t,x) = w(t,x) = W(t,x), \quad (t,x)\in [0,T]\times \mathbb{R} .
\end{equation*}
Hence, the determination of the pointwise optimal conditions confirms that $v^{\star }(\cdot )$ defined in \eqref{eq:v-star} is an optimal control strategy, and consequently $(\widetilde{Z}_{1}^{\star}, \widetilde{Z}_{1}^{\star})^{\top}(\cdot)$ defined in \eqref{eq:Ztilde-star} is a two-dimensional optimal control process that makes the backward component in \eqref{eq:qgFBSDE} an adapted process.

The uniqueness of the optimal solution is direct consequence of the uniqueness of the solution to the closed loop equation \eqref{eq:cle}. \qed

\subsection*{Proof of Corollary \ref{cor:Z-star}} \label{app:Z-star}

The result is obtained simply recalling \eqref{eq:Ztilde}, and applying the optimal backward process \eqref{eq:Ztilde-star} and the optimal trading strategy \eqref{eq:v-star}. \qed

\subsection*{Proof of Proposition \ref{prop:trading-vel}} \label{app:trading-vel}

The result follows from the explicit computation and taking into account the solution to the Riccati ODE for $a(\cdot)$ and the solution to the linear ODE derived for $\ell(T-\cdot)$, as specified in the proof of Lemma \ref{lemma:HJB-solution} at p. \pageref{app:HJB-solution}. \qed

\section{Dynamic risk measures} \label{app:risk-measures}

\begin{definition} \label{def:risk-measures}
A dynamic convex risk measure is a family of continuous semimartingales which maps, for any bounded stopping time $T$, a random variable $\xi (T) \in L_{\widetilde{\mathcal{F}}_{T}}^{2}\left(\Omega; \mathbb{R}\right)$ onto a process $\left\{\mathcal{R} \left(t,\xi (T)\right)\right\}_{t\in \left[ 0,T\right]}$ and satisfies the following axioms:

\smallskip

\noindent {\emph{Convexity:}} For any stopping time $S \leq T$, for any $\xi_{1}(T), \xi_{2}(T)$, for any $\alpha \in [0,1]$,
\begin{equation*}
\mathcal{R} \left(S,\alpha \xi_{1}(T) + (1-\alpha) \xi_{2}(T)\right) \leq \alpha \mathcal{R} \left(S,\alpha \xi_{1}(T)\right) \leq (1-\alpha) \mathcal{R}\left(S,\alpha \xi_{2}(T)\right) \quad \mathbb{P}\text{-a.s.};
\end{equation*}

\smallskip

\noindent {\emph{Decreasing monotonicity:}} For any stopping time $S
\leq T$, for any $\xi_{1}(T), \xi_{2}(T)$ such that $\xi_{1}(T) \geq \xi_{2}(T)$ $\mathbb{P}$-a.s., the operator is decreasing, i.e.,
\begin{equation*}
\mathcal{R}\left(S, \xi_{1}(T)\right) \leq \mathcal{R}\left(S, \xi_{2}(T)\right) \quad \mathbb{P}\text{-a.s.};
\end{equation*}

\smallskip

\noindent {\emph{Translation invariant}}: For any stopping time $S \leq T$, for any $\eta(S) \in \mathcal{F}_{S}$, for any $\xi(T)$,
\begin{equation*}
\mathcal{R}\left(S,\xi(T) + \eta(S)\right) = \mathcal{R}\left(S, \xi(T)\right) - \eta(S) \quad \mathbb{P}\text{-a.s.};
\end{equation*}

\smallskip

\noindent {\emph{Semigroup property} or \emph{Time consistency property}}: For any three bounded stopping time $S \leq T \leq U$, for any $\xi(U)$,
\begin{equation*}
\mathcal{R}\left(S, \xi(U)\right) = \mathcal{R}\left(S, -\mathcal{R}\left(T, \xi(U)\right)\right) \quad \mathbb{P}\text{-a.s.};
\end{equation*}

\smallskip

\noindent {\emph{Arbitrage free}}: For any stopping time $S\leq T$, for any $\xi_{1}(T), \xi_{2}(T)$ such that $\xi_{1}(T) \leq \xi_{2}(T)$ $\mathbb{P}\text{-a.s.}$,
\begin{equation*}
\mathcal{R}\left(S,\xi_{1}(T)\right) = \mathcal{R}\left(S,\xi_{2}(T)\right) \; \text{on} \; A_{S} =: \left\{S < T\right\} \; \Rightarrow \; \xi_{1}(T) = \xi_{2}(T) \quad \mathbb{P}\text{-a.s.} \; \text{on} \; A_{S}.
\end{equation*}
\end{definition}

In our framework, a generalized result regarding the strict relationship between dynamic convex risk measures and one-dimensional BSDEs is stated by the following proposition.

\begin{proposition} \label{th:risk-measure}
Let $\left(Z_{1}, Z_{2}\right)^{\top} := \{\left(Z_{1}, Z_{2}\right)^{\top} (t)\}_{t \in [0,T]}$ be the two-dimensional BSDE control process corresponding to the two-dimensional correlated Brownian motion $\left(B_{1}, B_{2}\right)$. If $g$ is a convex driver of a BSDE depending only on $\left(Z_{1}, Z_{2}\right)^{\top} \in \mathcal{H}_{\mathbb{F}}^{2}(0,T; \mathbb{R}^{2})$ then, for $\xi (T) \in L^{2}_{\mathcal{F}_{T}} (\Omega; \mathbb{R})$, the solution $\mathcal{R}(t,\xi(T))$ to:
\begin{equation*}
\begin{cases}
d\mathcal{R}(t, \xi (T)) = -g(t,Z_{1}(t), Z_{2}(t))dt + Z_{1}(t)dB_{1}(t) + Z_{2}(t)dB_{2}(t) \\
\mathcal{R}(T,\xi (T))=-\xi (T),
\end{cases}
\end{equation*}
characterizes a dynamic convex risk measure.
\end{proposition}

\begin{proof}
It follows straightforward from the result \cite[Theorem~3.21, pp.~125]{Barrieu-ElKaroui} and the comparison theorem in \cite[Theorem~5, p.~554]{Briand-Hu2}.
\end{proof}

Another simple example of $g$-conditional risk measure corresponding to a conventional mean-variance description of this risk return tradeoff is given by:
\begin{equation*}
g(t,Z(t)) = -\theta(t) Z(t)+\frac{1}{2}\left\Vert Z(t)\right\Vert^{2},
\end{equation*}
where $\theta$ can be interpreted as the correlation with the market.



\begin{thebibliography}{19}
\bibitem{Alfonsi-Schied}
{\sc Aur\'{e}lien Alfonsi} and {\sc Alexander Schied}.
{Optimal trade execution and absence of price manipulations in limit order book models},
\textit{SIAM Journal on Financial Mathematics},
\textbf{1},
pp. {490--522},
{2010}.

\bibitem{Almgren}
{\sc Robert Almgren}
{Optimal execution with nonlinear impact functions and trading-enhanced risk}.
\textit{Applied Mathematical Finance},
\textbf{10}(1),
pp. {1--18},
{2003}.

\bibitem{Almgren-Chriss}
{\sc Robert Almgren} and {\sc Neil Chriss},
{Optimal execution of portfolio transactions}.
\textit{Journal of Risk},
\textbf{3}(2),
pp.{5--39},
{2000}.

\bibitem{Bachelier}
{\sc Louis Bachelier},
{Th\'eorie de la sp\'eculation}.
\textit{Annales scientifiques de l'\'Ecole Normale Sup\'erieure},
s{\'e}rie 3, \textbf{17},
pp.{21--86},
{1900}.

\bibitem{Bank-Voss}
{\sc Peter Bank} and {\sc Moritz Vo{\ss}},
{Linear quadratic stochastic control problems with stochastic terminal constraint}.
\textit{SIAM Journal on Control and Optimization},
\textbf{56}(2),
pp.{672--699},
{2018}.

\bibitem{Barrieu-ElKaroui}
{\sc Pauline Barrieu} and {\sc Nicole El Karoui},
{Pricing, hedging, and optimally designing derivatives via minimization of risk measures}.
In R. Carmona (Ed.), \textit{Indifference pricing: Theory and applications},
pp. 77--146,
Princeton Series in Financial Engineering, Princeton University Press,
{2009}.

\bibitem{Bertsimas-Lo}
{\sc Dimitris Bertsimas} and {\sc Andrew W. Lo}.
{Optimal control of execution costs},
\textit{Journal of Financial Markets},
\textbf{1}(1),
pp.{1--50},
{1998}.

\bibitem{Bjork}
{\sc Tomas Bj\"{o}rk},
\textit{Arbitrage Theory in Continuous Time}.
{Second Edition},
{Oxford University Press},
{2009}.

\bibitem{Briand-Hu2}
{\sc Philippe Briand} and {\sc Ying Hu}.
{Quadratic BSDEs with convex generator and unbounded terminal conditions},
\textit{Probability Theory and Related Fields},
\textbf{141}(3-4),
pp.{543--567},
{2008}.

\bibitem{Brigo-DiGraziano}
{\sc Damiano Brigo} and {\sc Giuseppe Di Graziano}.
{Optimal execution comparison across risks and dynamics, with solutions for displaced diffusions},
\textit{Journal of Financial Engineering},
\textbf{1}(2),
pp.{1450018)
{2014}.

\bibitem{Bulthuis-et-al}
{\sc Brian Bulthuis, Julio Concha, Tim Leung} and {\sc Brian Ward},
{Optimal execution of limit and market orders with trade director, speed limiter, and fill uncertainty}.
\textit{International Journal of Financial Engineering},
\textbf{4}(02n03),
{2017}.


\bibitem{Carmona-Webster}
{\sc Ren{\'e} Carmona} and {\sc Kevin Webster},
{The self-financing equation in limit order book markets}.
\textit{Finance and Stochastics},
\textbf{23}(3),
pp.{729--759},
{2019}.

\bibitem{Cartea-et-al}
{\sc \'{A}lvaro Cartea, Sebastian Jaimungal} and {\sc Jos\'{e} Penalva},
\textit{Algorithmic and High-Frequency Trading}.
{Cambridge Univeristy Press},
{2015}.

\bibitem{Cartea-Jaimungal-SIAM}
{\sc \'{A}lvaro Cartea} and {\sc Sebastian Jaimungal},
{A closed-form execution strategy to target volume weighted average price}.
\textit{SIAM Journal on Financial Mathematics},
\textbf{7}(1),
pp.{760--785},
{2016}.

\bibitem{Xue}
{\sc Xue Cheng, Marina Di Giacinto} and {\sc Tai-Ho Wang},
{Optimal execution with uncertain order fills in Almgren-Chriss framework}.
\textit{Quantitative Finance},
\textbf{17}(1),
pp.{55--69},
{2017}.

\bibitem{Courtault-et-al}
{\sc Jean-Michel Courtault, Yuri Kabanov, Bernard Bru, Pierre Cr\'{e}pel, Isabelle Lebon} and {\sc Arnaud Le Marchand},
{Louis Bachelier on the centenary of Th\'{e}orie de la sp\'{e}culation}.
\textit{Mathematical Finance},
\textbf{10}(3),
pp.{339--353},
{2000}.

\bibitem{CurGatLil}
{\sc Gianbiagio Curato, Jim Gatheral} {\sc Fabrizio Lillo},
{Optimal execution with non-linear transient market impact},
\textit{Quantitative Finance}.
\textbf{17}(1),
pp.{41--54},
{2017}.

\bibitem{Cuoco-He-Isaenko}
{\sc Domenico Cuoco Hua He} and {\sc Sergey Isaenko},
{Optimal dynamic trading strategies with risk limits}.
\textit{Operations Research},
\textbf{56}(2),
pp.{358--368},
{2008}.

\bibitem{Engle-Ferstenberg}
{\sc Robert Engle} and {\sc Robert Ferstenberg},
{Execution risk},
\textit{Journal of Portfolio Management}.
\textbf{33}(2),
pp.{33--44},
{2007}.

\bibitem{Gatheral}
{\sc Jim Gatheral},
{No-dynamic-arbitrage and market impact}.
\textit{Quantitative Finance},
\textbf{10}(7),
pp.{749--759},
{2010}.

\bibitem{Gatheral-Schied}
{\sc Jim Gatheral} and {\sc Alexander Schied},
{Optimal trade execution under geometric Brownian motion in the Almgren and Chriss Framework}.
\textit{International Journal of Theoretical and Applied Finance},
\textbf{14}(3),
pp.{353--368},
{2011}.

\bibitem{Lee}
{\sc Kiseop Lee},
{Risk minimization under budget contraints}.
\textit{The Journal of Risk Finance},
\textbf{9}(1),
pp.{71--80},
{2008}.

\bibitem{Lin-Chen-Pena}
{\sc Qihang Lin, Xi Chen} and {\sc Javier Pe\~{n}a},
{A trade execution model under a composite dynamic coherent risk measure}.
\textit{Operations Research Letters},
\textbf{43}(1),
pp.{52--58},
{2015}.

\bibitem{Obizhaeva-Wang}
{\sc Anna A. Obizhaeva} and {\sc Jiang Wang},
{Optimal trading strategy and supply/demand dynamics}.
\textit{Journal ofFinancialMarkets},
\textbf{16}(1),
pp.{1--32},
{2013}.

\bibitem{Oksendal-Sulem}
{\sc Bernt {\O}ksendal} and {\sc Agn\`{e}s} Sulem},
{Maximum principles for optimal control of forward-backward stochastic differential equations with jumps}.
\textit{SIAM Journal on Control and Optimization},
\textbf{48}(5),
pp.{2945--2976},
{2009}.

\bibitem{Peng-1997}
{\sc Shige Peng},
{BSDE and stochastic optimization}.
{In J. Yan, S. Peng, S. Fang. and L. Wu},
\textit{Topics in Stochastic Analysis} (ch. 2), Science Press,
{1997}.

\bibitem{Peng-2004}
{\sc Shige Peng},
{Nonlinear expectations, nonlinear evaluations and risk measures}. {
In M. Frittelli and W. Runggaldier} (Eds.),
\textit{Stochastic Methods in Finance} (vol. 1856, pp. 165--253),
Lecture Notes in Mathematics,
{Springer-Verlag},
{2004}.

\bibitem{Perdoiu-Shakhet-Shreve}
{\sc Silviu Perdoiu, Gennady Shaikhet} and {\sc Shreve, Steven},
{Optimal execution in a general one-sided limit-order book}.
\textit{SIAM Journal on Financial Mathematics},
\textbf{2}(1),
pp.{183--212},
{2011}.

\bibitem{Vaes-et-Hauser-arxiv}
{\sc Julien Vaes} and {\sc Raphael Hauser},
{Optimal execution strategy under price and volume uncertainty}.
{Preprint}, available on ArXiv, {2018}.

\bibitem{Yong-Zhou}
{\sc Jongmin Yong} and {\sc Xun Yu Zhou},
\textit{Stochastic Controls: Hamiltonian Systems and HJB Equations}.
{Springer Verlag},
{1999}.
\end{thebibliography}
\end{document}